\documentclass[10pt,letterpaper]{article}
\pdfoutput=1

\usepackage[top=0.85in,left=1.75in,right=1.75in,footskip=0.75in]{geometry}

\usepackage{amsmath,amssymb}
\usepackage{changepage}
\usepackage[utf8x]{inputenc}
\usepackage{textcomp,marvosym}
\usepackage{cite}
\usepackage{nameref,hyperref}
\usepackage[right]{lineno}
\usepackage{microtype}
\DisableLigatures[f]{encoding = *, family = * }
\usepackage[table]{xcolor}
\usepackage{array}
\newcolumntype{+}{!{\vrule width 2pt}}
\newlength\savedwidth

\newcommand\thickhline{\noalign{\global\savedwidth\arrayrulewidth\global\arrayrulewidth 2pt}%
\hline
\noalign{\global\arrayrulewidth\savedwidth}}

\usepackage{setspace} 

\usepackage{lastpage,fancyhdr,graphicx}

\raggedright
\usepackage[aboveskip=1pt,labelfont=bf,labelsep=period,justification=raggedright,singlelinecheck=off]{caption}

\makeatletter
\renewcommand{\@biblabel}[1]{\quad#1.}
\makeatother

\date{}

\begin{document}
\vspace*{0.2in}

\begin{flushleft}
{\Large
\textbf\newline{Traveling pulse emerges from individuals coordinating their stop-and-go motion: a case study in sheep} 
}
\newline
\\
Manon Azaïs\textsuperscript{1},
Stéphane Blanco\textsuperscript{2},
Richard  Bon\textsuperscript{1},
Richard Fournier\textsuperscript{2},
Marie-Hélène Pillot\textsuperscript{1},
Jacques Gautrais\textsuperscript{1*}
\\
\bigskip
\textbf{1} Centre de Recherches sur la Cognition Animale (CRCA), Centre de Biologie Intégrative (CBI), Université de Toulouse; CNRS, UPS, France.
\\
\textbf{2} LaPlaCE, Université de Toulouse; CNRS, UPS, France.
\\
\bigskip

* jacques.gautrais@univ-tlse3.fr

\end{flushleft}
\section*{Abstract}
Monitoring small groups of sheep in spontaneous evolution in the field, we decipher behavioural rules that sheep follow at the individual scale in order to sustain collective motion.
Individuals alternate grazing mode at null speed and moving mode at walking speed, so cohesive motion stems from synchronising when they decide to switch between the two modes. 
We propose a model for the individual decision making process, based on switching rates between stopped / walking states that depend on behind / ahead locations and states of the others.
We parametrize this model from data.
Next, we translate this (microscopic) individual-based model into its density-flow (macroscopic) equations counterpart.
Numerical solving these equations display a traveling pulse propagating at constant speed even though each individual is at any moment either stopped or walking.
Considering the minimal model embedded in these equations, we derive analytically the steady shape of the pulse (sech square).
The parameters of the pulse (shape and speed) are expressed as functions of individual parameters.
This pulse emerges from the non linear coupling of start/stop individual decisions which compensate exactly for diffusion and promotes a steady ratio of walking / stopped individuals, which in turn determines the traveling speed of the pulse.
The system seems to converge to this pulse from any initial condition, and to recover the pulse after perturbation.
This gives a high robustness to this coordination mechanism.



\section*{Introduction}

Behavioural mechanisms driving collective motion in animals and chemotactic bacteria have raised a sustained interest over the last twenty years ~\cite{Sumpter2006,Eftimie2007,Tindall2008,Saragosti2010,Sumpter2010,Saragosti2011,Lopez2012a,Vicsek2012,Eftimie2012,Kuwayama2013,Carrillo2014,Pineda2015,Cavagna2016,Herbert-Read2016,Jiang2017}. 
Beyond attraction/repulsion basics, lots of studies have been devoted to understand how individuals coordinate their turns (velocity matching, in magnitude and direction), either considering a constant speed module ~\cite{Vicsek1995,Gautrais2012} or adaptive accelerations ~\cite{Katz2011,Tunstrom2013,Bialek2014}.
Data-based models have been proposed to understand mutual interactions within flocks and schools ~\cite{Hemelrijk2015,Hemelrijk2015a,Hemelrijk2014,Gautrais2012,Ballerini2008}, and how they translate into large-scale correlations and information propagation at the group scale ~\cite{Cavagna2016,Bialek2012,Attanasi2013a,Attanasi2015a,Calovi2015}.

Here, we focus on a specific kind of speed coordination, namely for terrestrial animals who display intermittent motion ~\cite{Kramer2001}: at any time, an individual is either stopped (null speed), or it is walking at a given constant speed. 
Such individual intermittent motion processes can combine into collective displays that could be poorly accounted for by continuous-speed models ~\cite{Ginelli2015,Rimer2017}.
In a stop-go process, the behavioural decision is about the delay before switching from stopped to walking, and back, depending on the relative position and moving states of the others.
In order to decipher the behavioural mechanisms at play so that intermittent-moving animals keep moving together, we studied small groups of sheep left alone grazing on their own on flat homogeneous pastures.
Our purpose is first to confirm that their decision making process in spontaneous condition in the field can be modelled by extending a previous model that accounted for their decision making process in manipulative condition. 
Our second purpose is to derive a macroscopic model based on this microscopic (individual) behavioral rules.

\section*{Biological background}
We followed the spontaneous evolution of groups of $N=2,3,4,8$ Merino sheep, introduced in fence-delimited square pens (80 x 80 m) planted in flat irrigated pastures (groups of $N=100$ sheep have also been monitored in the same series of experiments and have been the subject of a separate study, reported in a previous paper~\cite{Ginelli2015}). 

After habituation time, the groups adopted a collective behaviour alternating phases of quasi-static grazing and phases of head-up walking (see \nameref{S1_Movie} for an illustration with a group of three sheep).
There was a striking coordination of these phases among individuals, so that, most of time, a group is either found with all individuals grazing or all individuals walking.
The collective grazing phases are characterised by individuals hardly moving and keeping very close to each others (within the meter).
The collective walking phases are much shorter than grazing phases and can translocate the groups over several tens of meters.

\begin{figure}[!ht]
\centering
\includegraphics[width=.45\linewidth]{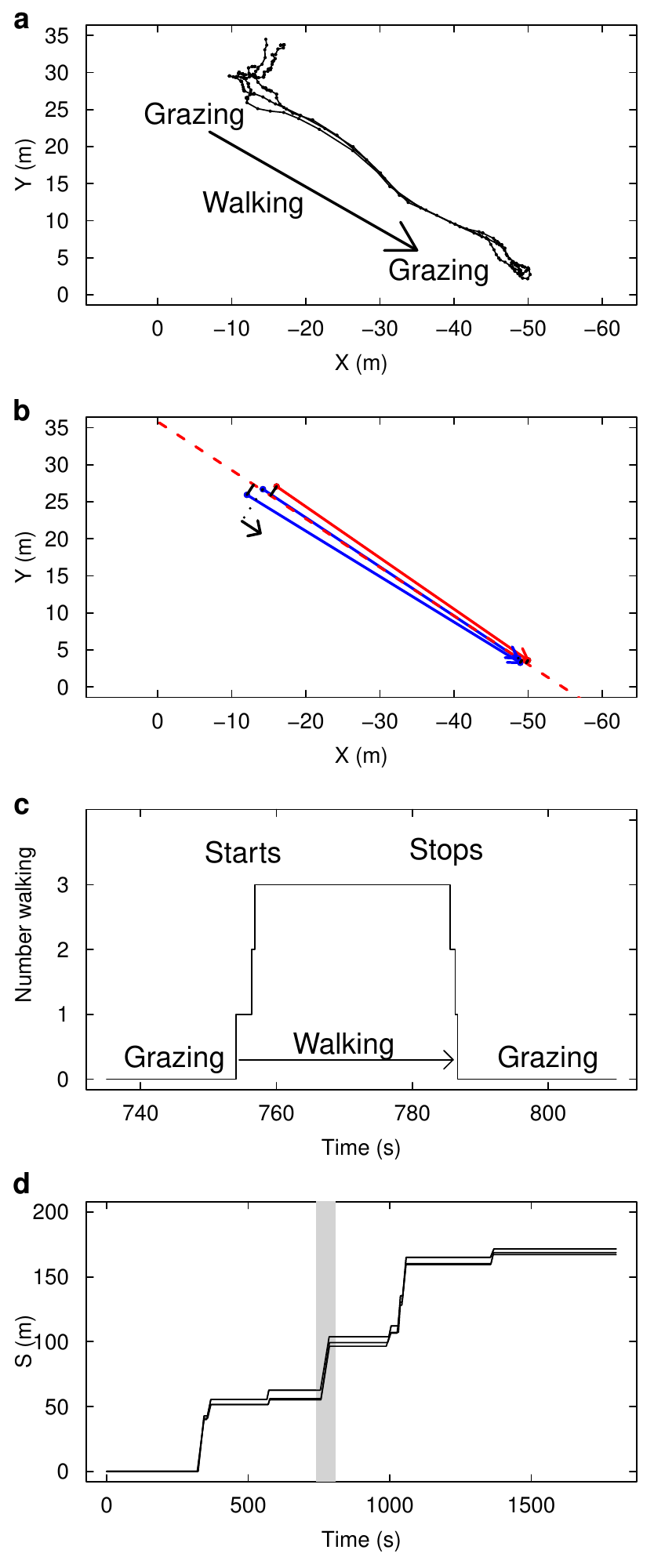}
\caption{{\bf Coordination of motion illustrated in one experimental group of 3 sheep.}
The position and behaviour of each individual is monitored every 1s during 1800 s. The collective behaviour can be categorised as periods of collective grazing (individuals are about motionless) interspersed by periods of collective walking (high speed motion).
(a) An extract of 70 s shows a typical event of collective transition from grazing to walking, leading to a spatial shift of the group to a new location where individuals resume grazing. 
(b) The data are idealised by binarizing individual speed (0,1) and the motion is projected in 1D along the axis of collective motion.
(c) The same event reported in time shows that the collective starts and stops are triggered by sheep synchronising their transition from grazing to walking (and back) within time windows by far shorter than typical duration of grazing / walking periods.
(d) Time-space representation of group evolution in 1D. Alternating synchronously grazing/walking/grazing over large time (1800 s) lead to a collective stop-and-go progression along the curvilinear abscissa $S$ of the group trajectory. The extracted event of 70 s is highlighted.}
\label{fig1}
\end{figure}

Starting from a collective state when all individuals are stopped, a collective grazing period ends when one individual spontaneously departs away from the group. 
This departure triggers a reaction in the others, who switch in turn to the walking state and follow the initiator.
The group then walk for a while until one of the sheep stops and resumes grazing, which in turn triggers the same behavioural switch in the others (Fig~\ref{fig1}a,b,c).
Since the characteristic duration of the grazing/moving periods are large compared to the duration of switching cascades (Fig~\ref{fig1}c), collective transition events (collective departures or collective stops) are well defined.

\clearpage

Regarding the directional process, we observed that the followers always adopted a bearing matching the initiator's (they \emph{followed} him, Fig~\ref{fig1}a), so that we will not address here the orientational decision, taking for granted that the initiator chooses a bearing, that the followers will systematically mimic.
We can thus consider the spatial progress of a group along the multi-segments trajectory of the group center of mass, indexing the individual positions by projecting their 2D positions onto the corresponding curvilinear abscissa along this group trajectory (Fig~\ref{fig1}b).
Doing this, the collective dynamics are idealised as individuals progressing in 1D towards positive abscissa (Fig~\ref{fig1}d), and the question becomes to understand the mechanisms  synchronising their switches from null-speed grazing to full-speed progression, and back (Fig~\ref{fig1}c).

In previous studies, we have proposed an individual-based model to explain the collective dynamics of group departures ~\cite{Petit2009a,Pillot2011}, and group stops ~\cite{Toulet2015} observed in a manipulative setup, using a remote control device to trigger the departure of a first (trained) individual ~\cite{Pillot2011,Toulet2015}.
In this model, individuals are in two possible states: stopped or moving (at speed $v$).
Their transitions from state to state are governed by a transition rate (probability switching state per unit time), which depends on the state configuration of the others.
In ~\cite{Pillot2011}, we only considered collective departures of naive individuals after the trained individual had departed.
We had found a double mimetic effect based on the state of the others: departed individuals tend to stimulate stopped individuals to switch to moving while still stopped individuals tends to inhibit it.
The higher the number of individuals that have already departed, the higher the rate to depart. The higher the number of individuals that are still stopped, the lower the rate to depart.
We proposed then a formal dependence of the stopped-to-moving (activation) switching rate $K_{_{A}}$, following:

\begin{equation}
K_{_{A}}(A,I) = \alpha \frac{A^{\beta}}{I^{\gamma}}
\label{eq:ibm1}
\end{equation}

where $A$ denotes the number of moving individuals (departed, active) and $I$ the number of stopped individuals (not departed, inactive).
We checked in a later study ~\cite{Toulet2015} that the same double mimetic effect can explain as well how moving-to-stopped (inactivation) switchings escalate in a group of moving individuals to reach a consensus to stop.

In those previous studies, only one event (collective departure or collective stop) was monitored at a time, a trained individual was used to trigger the collective events and the model was purely temporal.
To give account of groups behaviour in the present study, we start from the same model, to which we add two ingredients so that groups can chain multiple collective departures / collective stops as they meander spontaneously on the pasture. 

The first ingredient accounts for the spontaneous switching rates, to allow a first individual to depart from a stopped group and a first individual to stop in a moving group.

A second ingredient is needed to introduce spatial effects.
In the present setup (small groups on open pastures), it was obvious that each sheep can monitor every other one, so we do not introduce limited range of interaction (this point is discussed further in the Discussion), neither metric nor topologic ~\cite{Ballerini2008}.
As a proxy for the relevant information in sheep decisions, we consider only relative positions along the 1-dimensional group trajectory, so that one individual can make a difference between individuals ahead of him and individuals behind him.
The states configuration of the others around can then be split into four pools: the individuals behind him that are stopped  $I^{-}$, the ones behind him that are moving $A^{-}$, the ones ahead that are stopped $I^{+}$ and the ones ahead that are moving $A^{+}$.

For the stopped-to-moving switching rates $K_{_{A}}$ (activation), the double mimetic effects become:

\begin{equation}
\begin{array}{ll}
K_{_{A}}(A^{-},I^{-},A^{+},I^{+}) & = \mu_{_{A}} + \alpha_{_{A}} \frac{\left[A^{+}\right]^{\beta_{_{A}}}}{\left[A^{-}+I^{-}+I^{+}\right]^{\gamma_{_{A}}}} \\
\ \\
& = \mu_{_{A}} + \alpha_{_{A}} \frac{\left[A^{+}\right]^{\beta_{_{A}}}}{\left[N-A^{+}\right]^{\gamma_{_{A}}}}
\end{array}
\label{eq:ibmact}
\end{equation}

and for the moving-to-stopped switching rates:

\begin{equation}
\begin{array}{ll}
K_{_{I}}(A^{-},I^{-},A^{+},I^{+}) &= \mu_{_{I}} + \alpha_{_{I}} \frac{\left[I^{-}\right]^{\beta_{_{I}}}}{\left[A^{+}+A^{-}+I^{+}\right]^{\gamma_{_{I}}}} \\
\ \\
&= \mu_{_{I}} + \alpha_{_{I}} \frac{\left[I^{-}\right]^{\beta_{_{I}}}}{\left[N-I^{-}\right]^{\gamma_{_{I}}}}
\end{array}
\label{eq:ibminact}
\end{equation}

where we have considered that only neighbours ahead and moving, $A^{+}$, are stimulating switches to motion (the others inhibiting it) and only stopped neighbours behind, $I^{-}$, are stimulating stopping decision of a moving animal (the others inhibiting it). This modeling choice is discussed further in the Discussion.
In absence of stimulating individuals, the rates reduce to the spontaneous switching rates $\mu_{_{A}}$ and $\mu_{_{I}}$.

\subsection*{Parameters estimation}
In order to estimate the parameters for the stimulated part ($\alpha_{\mathrm{\bullet}},\beta_{\mathrm{\bullet}},\gamma_{\mathrm{\bullet}}$), we collected all collective transition events from  1800-s movie sequences, combining group sizes to disentangle the two mimetic effects, as in ~\cite{Pillot2011,Toulet2015} (Table~\ref{table1}). 

\bigskip

\begin{table}[!h]
\centering
\caption{\bf{Individual parameters for the double mimetic effect.} }
\begin{tabular}{|l+r|r|r|r|}
\hline
Parameter & Stopped-to-Moving & Kept & Moving-to-Stopped & Kept \\ \thickhline
$\alpha \ (s^{-1})$ & 0.32 [0.25;0.41] & 0.3 &0.42 [0.33;0.54] & 0.4\\ \hline
$\beta$ & 0.61 [0.44;0.78] & 0.6 &0.48 [0.31;0.65] & 0.5\\ \hline
$\gamma$ & 0.71 [0.53;0.87] & 0.7 &0.54 [0.36;0.71] & 0.5\\ \hline
\end{tabular}
\begin{flushleft} Table notes Mean estimates and 95\% CI are given for each parameter and for both kinds of transition. In Kept columns are reported the mean estimates rounded at the first decimal, which was retained in the model.\end{flushleft}
\label{table1}
\end{table}
 
\bigskip

\begin{figure}[!ht]
\centering
\includegraphics[width=0.7\linewidth]{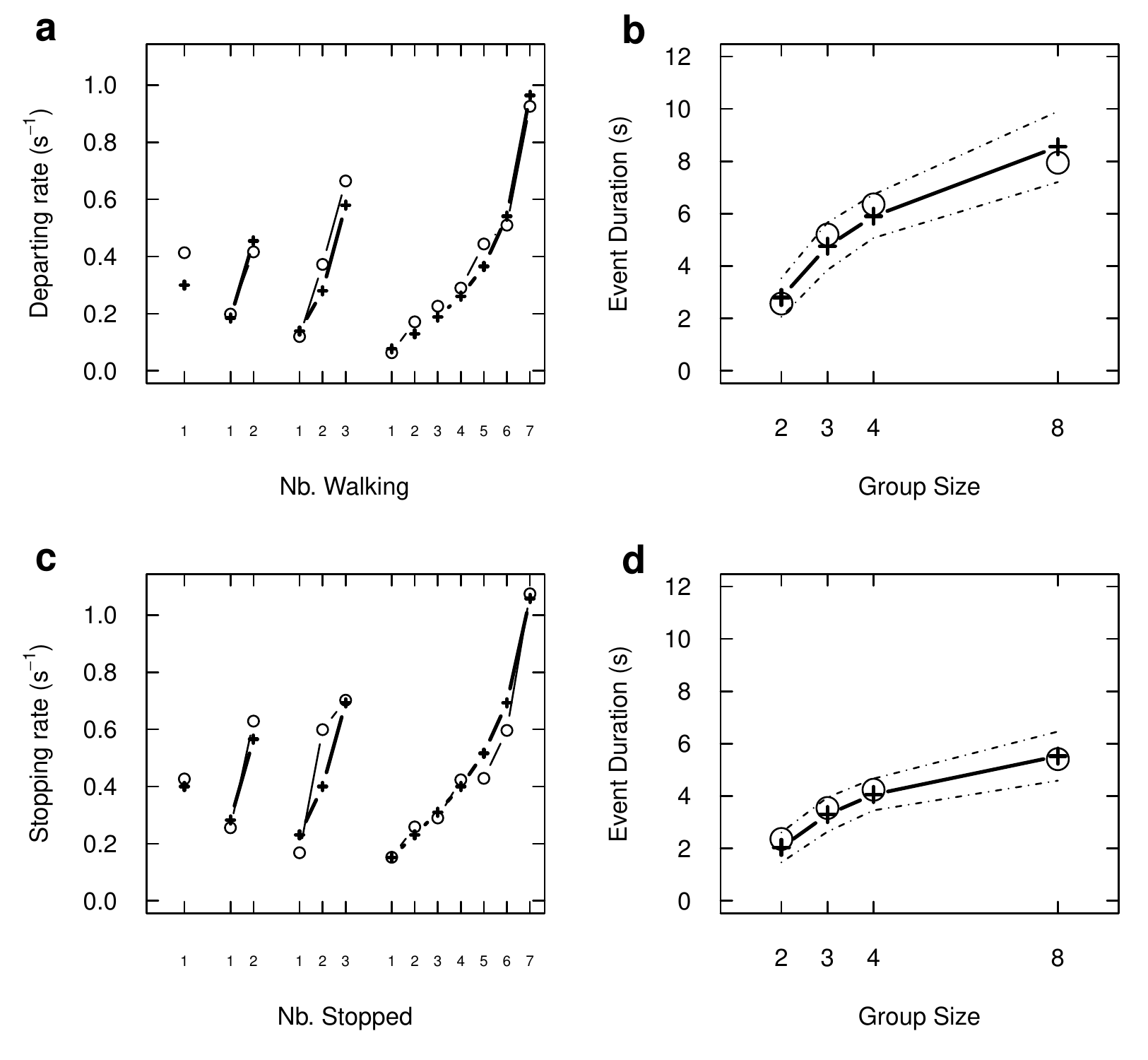}
\caption{{\bf Mimetic amplification governs individual transition rates.} 
(a) For each group size (2,3,4,8), we report the individual transition rate from stopped to moving as a function of the number of individuals moving ahead. This rate increases, indicating that individuals moving ahead have a positive feedback effect upon the propensity to follow them (stimulating effect). Note that, for a given number of departed individuals (e.g. 1), the rate decreases with group size, indicating an inhibitory effect of other individuals. Open circles: data, crosses: fitted rates.
(b) For each group size, simulating collective departures using the fitted transition rates yields a correct prediction of the average event duration, from first start to last start, as a function of group size (10000 simulated events, dotted lines indicate 95\% CI of the mean).
(c) Same kind of data and fitted rates, but for the moving-to-stopped rates. While en route, stopping rates are positively enhanced by the number of individuals stopped behind (stimulating effect), together with a inhibitory effect of the others.
(d) Events simulations also confirm that the fitted stopping rates yield correct predicted average event duration, from first stop to last stop, as a function of group size.
Fitted parameters are indicated in Table \ref{table1}.}
\label{fig2}
\end{figure}

This functional dependence fitted with the experimental rates as nicely as in our previous studies (Fig~\ref{fig2}a,c), and correctly predicted as well the duration of events depending on group size (Fig~\ref{fig2}b,d).

The spontaneous rate of switching to the stopped state $\mu_{I}$ was straightforwardly retrieved from collective moves duration, and we found that it depends on the group size $N$, following:

\begin{equation}
\mu_{I} = \mu^{*}_{I}/N
\label{eq:mun}
\end{equation}
with $\mu^{*}_{I}= 0.08 \ s^{-1}$.

\begin{figure}[!ht]
\centering
\includegraphics[width=0.7\linewidth]{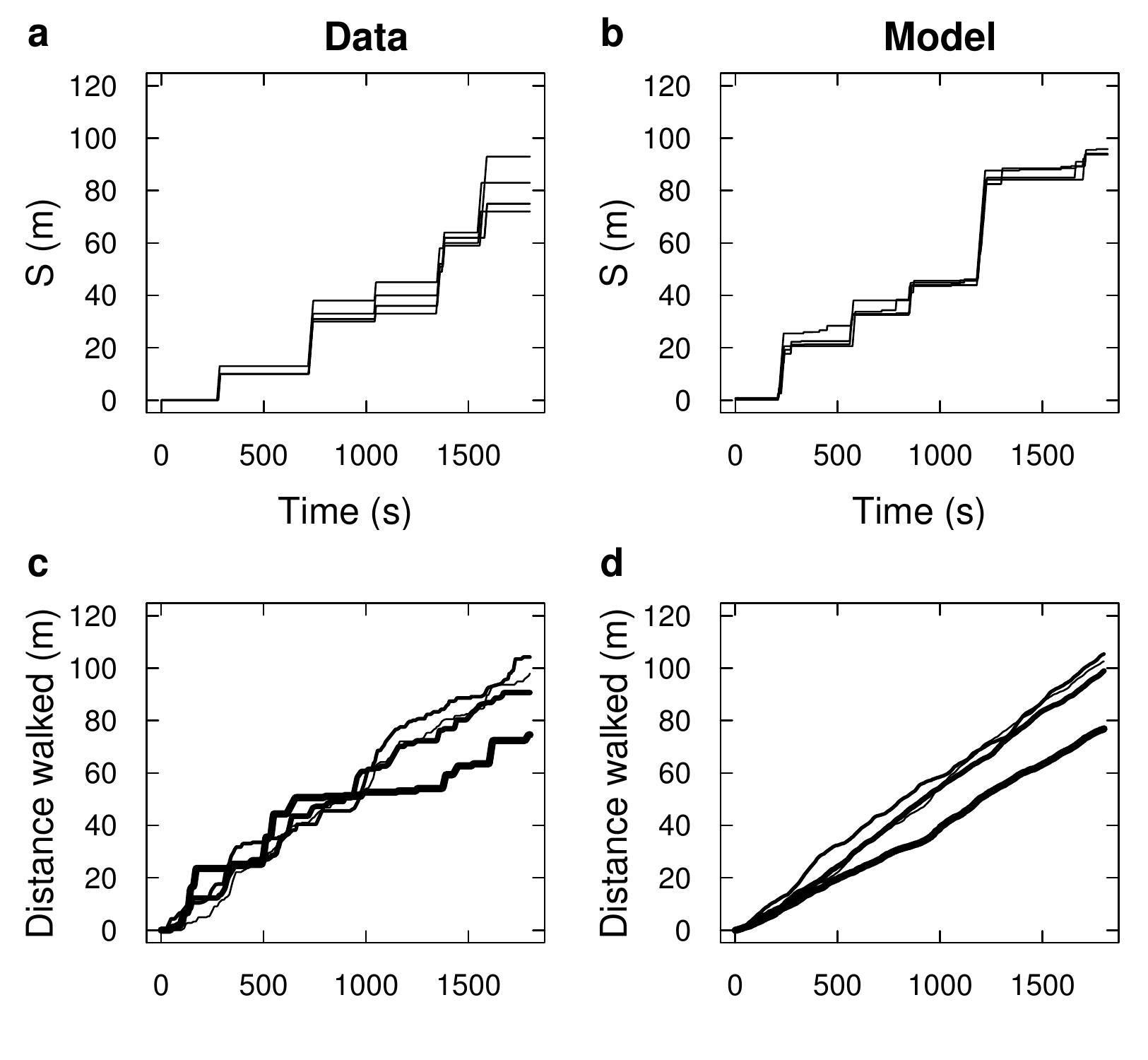}
\caption{{\bf Individual-Based Model (IBM) prediction for 1D-propagation.} 
(a) The evolution of one experimental group of 4 sheep is reported for illustration (same time-space representation as Fig~1d).
(b) A typical evolution of a simulated 4-sheep group is reported for visual comparison with (a). This evolution is one stochastic realisation of the IBM, computed with an exact Monte Carlo (Gillespie algorithm was used), using the fitted rates and $\mu_{A} = 0.0055 \ (s^{-1})$.
(c) Average distance walked by experimental groups over 1800 s, for each group size separately (thiner line: N=2, thicker line: N=8).
(d) Corresponding IBM predictions, averaging over 100 simulations per group size, like the one reported in (b).}
\label{fig3}
\end{figure}

The spontaneous rate of switching to the walking state was practically impossible to estimate from data because small grazing moves and actual departures as an initiator were too difficult to discriminate.
A reasonable estimate is however $\mu_{A} = 0.0055 \ (s^{-1})$, corresponding to a mean time of 3 minutes before next spontaneous departure.
This estimate yields Monte Carlo realisations that singly compare favourably with experimental alternation of stopped / moving periods (Fig~\ref{fig3}b vs. \ref{fig3}a), and on average with the distances covered by the groups over 1800 s (Fig~\ref{fig3}d vs. \ref{fig3}c).

We note that this value is inevitably \emph{ad hoc} for the present experimental conditions, and may vary a lot depending on the available resources, the animals' physiological state, the day hour, and the seasons. 
Still, the cohesion of collective departures and collective stops would poorly depend on $\mu_{A}$, provided it remains low in front of the stimulating ingredient, which is well the case here ($\mu_{A} \ll \alpha_{A}$). 
Indeed, stochastic simulations show that this low value guarantees a sustained cohesion (see \nameref{S2_Movie}).
In the opposite, for spontaneous rates dominant over coupling ($\mu_{A} \gg \alpha_{A}$), the groups would consistently disperse (see \nameref{S3_Movie}).

Overall, we advocate that the proposed model give a good account of the behavioural mechanism driving the sustained groups cohesion when small groups of sheep are left pasturing on their own. 

\section*{Deriving the macroscopic model}

We turn now to the theoretical study of this collective motion emerging from $N$ interacting stop-and-go individuals, should they be sheep or any other entities.
The intriguing feature of this kind of collective motion is that \emph{groups can be seen to progress as a whole at some fraction of the individual speed even though each individual is either stopped or moving at full speed at any time}.
This is well illustrated in the \nameref{S4_Movie}, with a group of N=32 sheep.

The interesting observables at collective level are then the collective speed at which groups propagate on the one hand, and how cohesive they remain in time on the other hand.
To better understand how these collective observables emerge from individual behaviours, we translate the individual-based model exposed above into densities equations.
In this aim, we provisionally admit that the model extends as it is to large groups.
We first translate the model into the corresponding Boltzmann equations, check for finite size effects and then derive the density-flow equations.

\subsection*{Boltzmann-like (kinetic) equations}
Let $A(x,t)$ and $I(x,t)$ denote respectively the density of active (moving) and inactive (stopped) sheep at location $x$ at time $t$. 
They evolve according to:

\begin{equation}
\left\{ \begin{array}{lll}
\partial_{t}I(x,t)&=&-K_{A}(x,t)I(x,t)+K_{I}(x,t)A(x,t)\vspace{0.2cm}\\
\partial_{t}A(x,t)+v\partial_{x}A(x,t)&=&+K_{A}(x,t)I(x,t)-K_{I}(x,t)A(x,t)  
\end{array} \right.
\label{eq:boltz1}
\end{equation}

where $K_{A}(x,t)$ and $K_{I}(x,t)$ are respectively the conversion rates from stopped-to-moving (activation) and moving-to-stopped (inactivation) at location $x$ at time $t$, which depend on $A$ and $I$ according to:

\begin{equation}
\left\{ \begin{array}{lll}
K_{A}(x,t) &= \mu_{_{A}} 
 & + \alpha_{_{A}}{\left[  \int^{\infty}_{x} A(u,t)du  \right]}^{\beta_{A}} {\left[ N - \int^{\infty}_{x} A(u,t)du  \right]}^{-\gamma_{A}} \vspace{0.2cm}\\
K_{I}(x,t) &= \mu_{_{I}} 
 & + \alpha_{_{I}}{\left[  \int^{x}_{-\infty} I(u,t)du  \right]}^{\beta_{I}} {\left[ N - \int^{x}_{-\infty} I(u,t)du  \right]}^{-\gamma_{I}} 
\end{array} \right.
\label{eq:boltz2}
\end{equation}

with $N=\int^{\infty}_{-\infty} A(u,t)+I(u,t)du$ is the total amount of sheep (which is conserved in time), and parameters are those given in the individual-based model.
This description in density is the direct translation of the individual model expressions (in the limit of continuum theory).

\begin{figure}[!ht]
\centering
\includegraphics[width=0.8\linewidth]{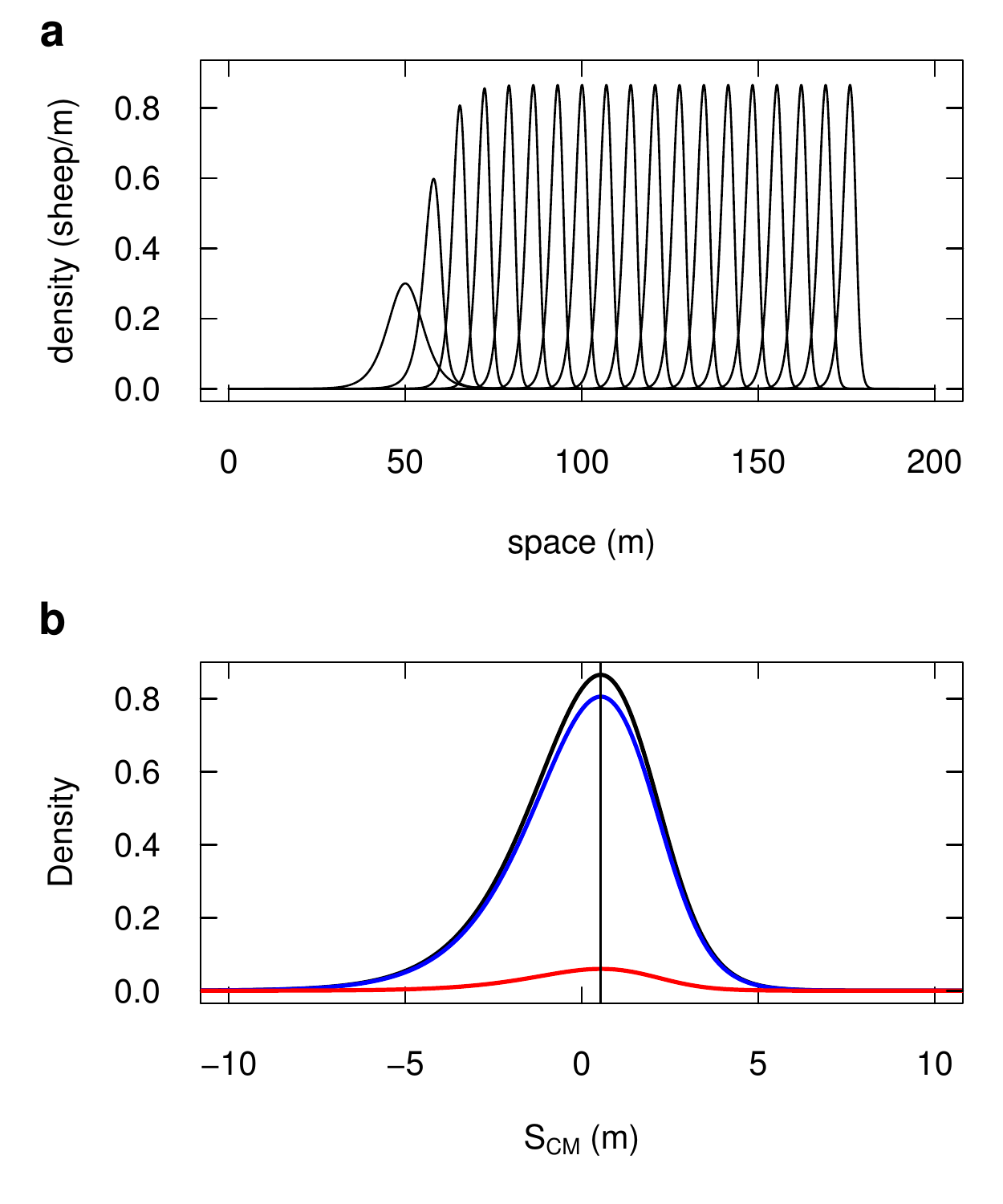}
\caption{{\bf  Density Model prediction for 1D-propagation.}
(a) The predicted evolution of a group of 4 sheep is the formation of a traveling pulse, which travels at constant speed, and with no dispersion (the leftmost profile is the initial condition, the rightmost profile is at time 1800 s, intermediate profiles are every 100 s ; numerical solution of Eqs.~\ref{eq:boltz1}-\ref{eq:boltz2}, with $\Delta t=10^{-2}$ s).
(b) The stabilised profile (black line, at time 1800 s) is zoomed out to show the density distribution around the group center of mass (0 abscissa). 
It displays a slight asymmetry (vertical line through the distribution peak for visual guidance), with an excess of density in the left tail (at the rear of the group). The underlying densities of stopped (blue) and moving (red) appear homogeneously proportional to the total.}
\label{fig4}
\end{figure}

Numerical resolution of Eqs~\ref{eq:boltz1}-\ref{eq:boltz2} shows that the density propagates as a cohesive traveling pulse (a solitary wave) \cite{Eftimie2007a,Kerner1994,Purwins2005,Purwins2010}, with no dispersal (Fig \ref{fig4}).
In contrast, the spontaneous evolution in absence of coupling ($\alpha_{A}=\alpha_{I}=0$) would propagate like a linear advection-diffusion system (see Fig \ref{S1_Fig}).
The system appears to converge to this solution for various initial conditions (see \nameref{S5_Movie}).
We have not found initial conditions that would lead to another solution, and we do not see actually which other solution there could be. 

\subsection*{Group size effects}

In the classical view (e.g. in gas and fluid mechanics), the spatiotemporal density equations represent the evolution of a continuous mass. 
That would correspond here to a group of infinite size. 
In linear systems, it can also represent straightforwardly the average statistic of presence over an infinite number of stochastic replicates, in which case density becomes probability density of presence in space-time.
In the present case, we deal with small groups governed by non linear mechanisms, so we had to ensure how well our Boltzmann equations reflect the behaviour of groups.
For this, we compared its predictions to individual-based model simulations.

In absence of coupling ($\alpha_{A}=\alpha_{I}=0$), the system behaves as an advection-diffusion process (which is linear), and we found a perfect convergence of Monte Carlo simulations of the Individual-Based Model and numerical resolution of Eqs~\ref{eq:boltz1}-\ref{eq:boltz2}, as expected (see Fig \ref{S2_Fig}).

Since groups can progress at quite different average speeds (see \nameref{S2_Movie}), averaging plainly individual presence over replications would yield the same kind of advection-diffusion pattern (because of the dispersion of the groups centres of mass, namely the inter-group variance), and it would not capture the cohesion within each group (intra-group variance).
To extricate inter-group variance from intra-group variance, we thus retained two separate statistics: the average speed of groups, and the internal dispersion within groups.
The internal dispersion is computed taking individual abscissa relative to the center of mass of the group they belong to, group by group, so that we superimpose statistics of presence centred around the center of mass.

\begin{figure}[!ht]
	\centering
        \includegraphics[width=0.8\linewidth]{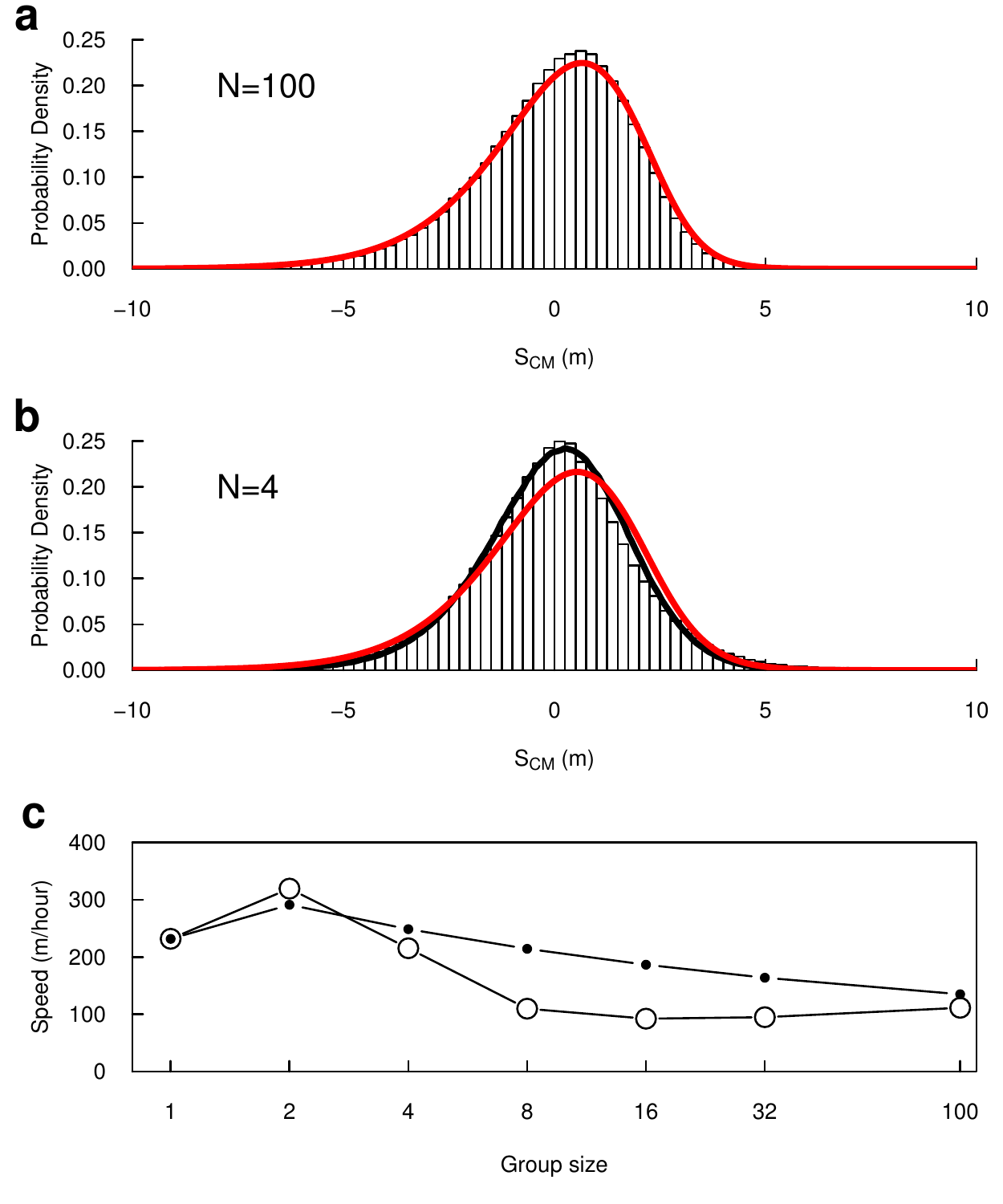}
        \caption{{\bf Density Model predictions vs. IBM predictions.} 
(a) Histogram: statistics of presence around the center of mass of the group for N=100, predicted from 300 IBM realisations ; red line: numerical solution of Eqs.~\ref{eq:boltz1}-\ref{eq:boltz2} ($\Delta t=10^{-2}$ s).
(b) Histogram: statistics of presence around the center of mass of the group for N=4, predicted from $10^{6}$ IBM realisations; red line: numerical solution of Eqs.~\ref{eq:boltz1}-\ref{eq:boltz2} ($\Delta t=10^{-2}$ s); black line: statistics of presence around the center of mass for groups of 4 positions sampled from the red curve, and applying the same procedure than the one used to obtain the histogram from IBM realisations ($10^{6}$ samples).
(c) Predicted propagation speed depending on group size. Open dots: IBM predictions (error bars lie within the point size). Black dots: Density Model predictions.\\
}
\label{fig5}
\end{figure}

Considering large groups (N=100), the numerical solution of Eqs~\ref{eq:boltz1}-\ref{eq:boltz2} fit nicely with averages over Individual-Based Model (IBM) stochastic realisations, both for the internal dispersion (Fig~\ref{fig5}a) and for the average speed (Fig~\ref{fig5}c, N=100).

Considering smaller groups (from N=32 down to N=2), we observe that the IBM estimation of internal dispersion tends to appear more symmetrical than the predicted continuous profile (Fig~\ref{fig5}b, for N=4). 
However, this finite size effect is only due to the observable itself because the center of mass computed from individuals' locations tend to be more stochastic in small groups. 
Indeed, if we sample groups of 4 positions from the continuous profile (Fig~\ref{fig5}b, red curve), and compute the same statistic as we do from IBM predictions, both fit very well (Fig~\ref{fig5}b, black curve).

In contrast, we observe a clear finite size effect regarding the average propagation speed (Fig~\ref{fig5}c): the IBM realisations with small groups show slower propagation than predicted by Eqs~\ref{eq:boltz1}-\ref{eq:boltz2}, albeit both follow the same trend.
This discrepancy will deserve further investigations in the future.

All in all, the Boltzmann equation captures the essential behaviour of groups as small as N=4, and represents one group progressing at constant speed, and keeping its density unchanged at long time: on average, groups behave as a traveling pulse.
Boltzmann equation could then be used directly to numerically explore properties of groups propagation depending on individual parameters, especially pulse shape and extension.
In the next section, we start from it  to progress toward an analytical solution.

\subsection*{Minimal model and analytical solution}
In an approach by minimal model, we are interested to realise the very essence of the coupling between individual and collective scales that sustains the propagation of the pulse in the steady regime.
To this end, we simplify as far as possible the model presented above by setting $\beta_{A}=\beta_{I}=1$, and neglecting inhibitory effects: $\gamma_{A}=\gamma_{I}=0$. 
Doing this, we keep only the two essential components: the spontaneous switch of speed (driven by $\mu_{A}$ and $\mu_{I}$), and the stimulating effect of the others (driven by $\alpha_{A}$ and $\alpha_{I}$). 

To derive an analytical solution, we first translate the Boltzmann equations above into the corresponding ``macroscopic'' density-flow equations.
We selected eventually two variables to describe this evolution. 
The dispersion can be described by the sum density of sheep $\eta(x,t)$ (moving and stopped) at location $x$ at time $t$:

\begin{equation}
\eta(x,t) = A(x,t) + I(x,t)
\end{equation}

and the collective speed can be described by the moving fraction $\beta(x,t)$ at location $x$ at time $t$:

\begin{equation}
\beta(x,t) = \frac{A(x,t)}{A(x,t) + I(x,t)} = \frac{A(x,t)}{\eta(x,t)}
\end{equation}

the collective speed being $v\ \beta(x,t)$, where $v$ is the speed at which a walking individual walks.

Next we consider the traveling pulse at the steady regime, and write the equation governing its density profile $n(y)$ in a moving frame anchored to the pulse peak ($y$ indexing abscissa in the moving frame).
This profile obeys:

\begin{equation}
(n')^{2} - nn''-\frac{\alpha_{A}  + \alpha_{I}}{v}n^{3}=0
\label{eq:steady7}
\end{equation}

where prime denotes regular derivative with respect to $y$.

A solution to Eq.\ref{eq:steady7} is given by:

\begin{equation}
n(y)= \frac{1}{2}\mathrm{N}\gamma \mathrm{sech}^{2}\left(\gamma y\right)\ \mathrm{with}\ \gamma = \frac{\mathrm{N}(\alpha_{A}+\alpha_{I})}{4v}
\label{eq:steady12}
\end{equation}

Full details for the analytical solution in the steady regime for the minimal model are given in \nameref{S1_Appendix}.

This solution well recovers the numerical predictions of the Boltzmann expression Eqs~\ref{eq:boltz1}-\ref{eq:boltz2}, for different parameters $\alpha_{\mathrm{\bullet}}$ (see see \nameref{S6_Movie} in which we have superimposed this analytical solution in red upon the numerical prediction in black).

The full expression for the pulse profile in the field frame and the associated propagation speed are:

\begin{equation}
\left\{ \begin{array}{lll}
\eta^{s}(x,t) &= \frac{\mathrm{N}}{2}\frac{\mathrm{N}(\alpha_{A}+\alpha_{I})}{4v} \mathrm{sech}^{2}\left(\frac{\mathrm{N}(\alpha_{A}+\alpha_{I})}{4v} \ (x-b_{s}^{*}vt)\right)\vspace{0.3cm}\\

\beta^{s}(x,t) &= b_{s}^{*} = \frac{\left( \frac{N}{2}(\alpha_{A}-\alpha_{I}) - \mu_{A}-\mu_{I} \right) +\sqrt{\left( \frac{N}{2}(\alpha_{A}-\alpha_{I}) - \mu_{A}-\mu_{I} \right)^{2} +4\mu_{A}\frac{N}{2}(\alpha_{A}-\alpha_{I})}}{N(\alpha_{A}-\alpha_{I})}
\end{array} \right.
\label{eq:steady21}
\end{equation}

\bigskip
when $\alpha_{A}\neq\alpha_{I}$.

In the symmetrical case where $\alpha_{A}=\alpha_{I}$, the steady state solution simplifies to:

\begin{equation}
\left\{ \begin{array}{lll}
\eta^{s}(x,t) &= \frac{N}{2}\frac{N\alpha}{2v} \ {\mathrm{sech}}^{2}(\frac{N\alpha}{2v} \ (x-b_{s}^{*}vt))\vspace{0.3cm}\\

\beta^{s}(x,t) &= b_{s}^{*} = \mu_{A}/(\mu_{A}+\mu_{I})
\end{array} \right.
\label{eq:steady22}
\end{equation}

The shape of the pulse depends on individual speed $v$, reaction terms $\alpha_{A}$ and $\alpha_{I}$ and group size $N$, while the propagation speed of the pulse  also depends on the spontaneous switching rates $\mu_{A}$ and $\mu_{I}$. 
Sensitivity of the latter to some parameters is illustrated in next section.

\subsection*{Sensitivity of the moving fraction to parameters}

The collective behaviour of first interest is the mean speed at which groups propagate, and it is a direct reflect of the moving fraction, given by Eq~\ref{eq:steady21}.
In the general case, this moving fraction depends upon two kinds of parameters:
\begin{enumerate}
\item The rates of spontaneous switching $\mu_{A}$ and $\mu_{I}$: in terms of individual behaviour, these rates govern the propensity for an individual to be the first to depart from a stopped group, respectively the propensity for an individual to be the first to stop in a moving group.
\item The imitation rates $\alpha_{A}$ and $\alpha_{I}$: in terms of individual behaviour, these rates govern the propensity for a stopped individual to imitate departing individuals, respectively the propensity for a moving individual to imitate stopping individuals.
\end{enumerate}

\begin{figure}[!ht]
	\centering
        \includegraphics[width=0.8\linewidth]{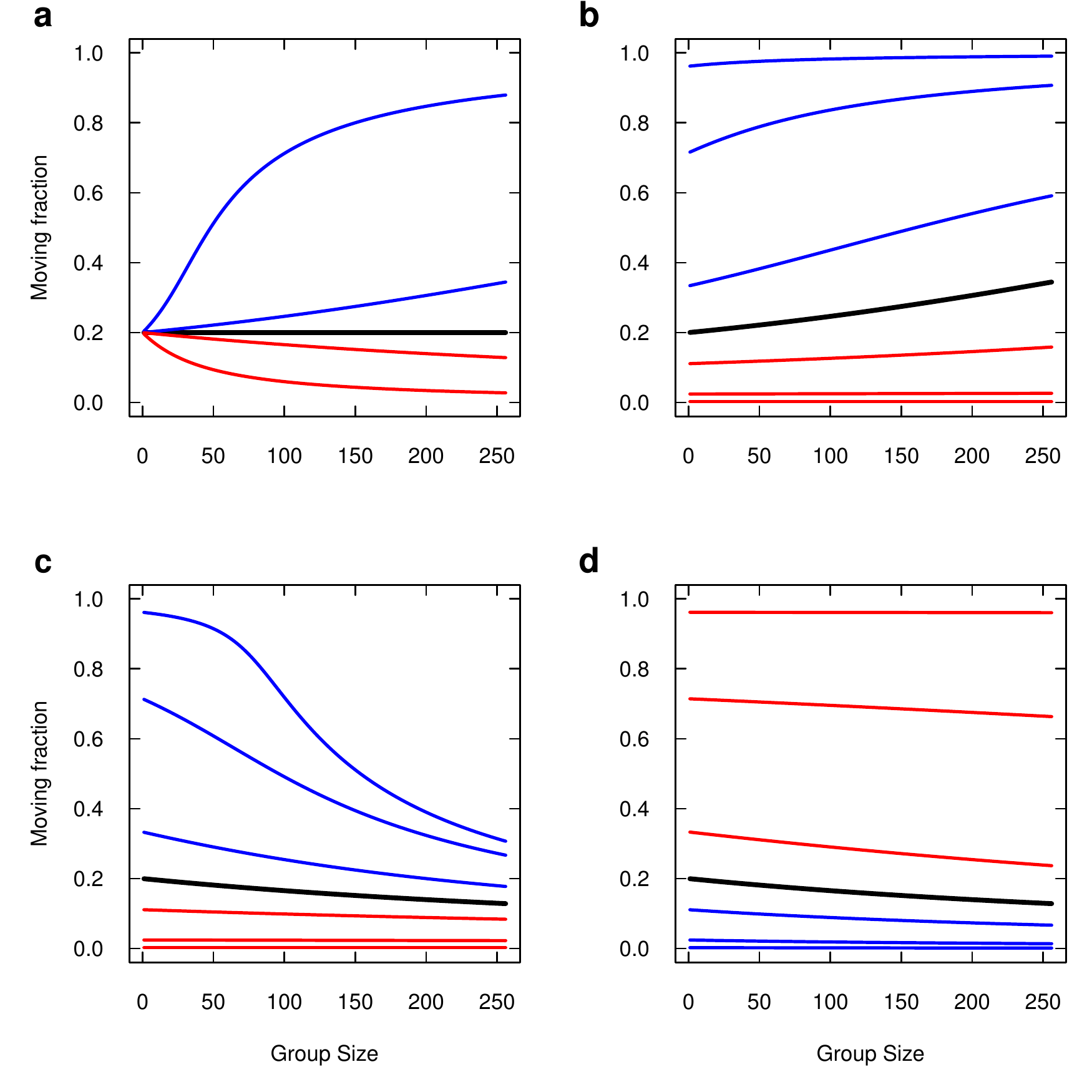}
        \caption{{\bf Moving fraction as a function of group size, varying individual parameters.} 
(a) The symmetrical imitations case is reported in black, with $\alpha_{A}=\alpha_{I}=\alpha=0.5$, $\mu_{A}^{*}=0.02$ and $\mu_{I}^{*}=0.08$. 
Blue curves: $\alpha_{A}=1.001\alpha$ and $\alpha_{A}=1.01\alpha$ keeping $\alpha_{I}=\alpha$, and red curves: same variations for $\alpha_{I}$ keeping $\alpha_{A}=\alpha$ .
(b) Setting $\alpha_{A}=1.001\ \alpha$, $\mu_{I}$ is varied from lower (blue) to higher (red) values than $\mu_{I}^{*}$ (black). Variations correspond respectively to division or multiplication by 2, 10 and 100.
(c) Setting $\alpha_{I}=1.001\alpha$, $\mu_{I}$ is similarly varied from lower (blue) to higher (red) values than $\mu_{I}^{*}$ (black).
(d) Setting $\alpha_{I}=1.001\alpha$, $\mu_{A}$ is similarly varied from lower (blue) to higher (red) values than $\mu_{A}^{*}$ (black).
}
\label{fig6}
\end{figure}

Fig~\ref{fig6} reports the moving fraction as a function of group size, varying the parameters.

As mentioned above, in the case of symmetrical imitations, the moving fraction $b_{s}^{*}$ appears to depend only on the rates of spontaneous switching $\mu_{A}$ and $\mu_{I}$ and in this particular case, it would not depend on the group size $N$ (Fig~\ref{fig6}a, black curve).
In the asymmetric cases, the moving fraction also depends on imitation rates and on group size.
Promoting departure imitation over stopping imitation ($\alpha_{A}>\alpha_{I}$), even by the slightest amount (Fig~\ref{fig6}a, blue curves, $\alpha_{A}=1.001\ \alpha_{I}$ and $\alpha_{A}=1.01\ \alpha_{I}$) lead larger groups to display higher and higher moving fractions (tends to 1 for large groups).
Conversely, promoting stopping imitation (Fig~\ref{fig6}a, red curves) lead larger groups to lower and lower moving fractions (tends to 0 for large groups). 

The trend to full moving fractions for promoted departure imitation depends on the spontaneous stopping rate $\mu_{I}$ (Fig~\ref{fig6}b). 
For very large values of $\mu_{I}$ (low moving fraction, Fig~\ref{fig6}b, red curves), this trend is very slow and might be negligible. 
At the other end of the scope, very low $\mu_{I}$ values would promote a high moving fraction even for smallest group (Fig~\ref{fig6}b, blue curves) so that the trend is also saturated. In between, $\mu_{I}$ has a sensible effect upon the trend.
The trend to null moving fractions for promoted stopping imitation is more affected by $\mu_{I}$ (Fig~\ref{fig6}c) than by $\mu_{A}$ (Fig~\ref{fig6}d), especially when it is low (Fig~\ref{fig6}c, blue curves).

Overall, the moving fraction depending on group size is especially sensitive to slight promotion of imitation rates, but also to spontaneous stopping rate when the latter is low.

\bigskip

As a variant of the model, our data suggest that the spontaneous stopping rate is regulated by the group size $N$ (Eq.\ref{eq:mun}) such that walking individuals in large groups tend to spontaneously stop less often.
In such a case, the moving fraction in the symmetrical imitation case, should be rewritten as:

\begin{equation}
b_{s}^{*} = \frac{\mu_{A}}{\mu_{A}+(\mu_{I}/N)}
\label{eq:steady23}
\end{equation}

\begin{figure}[!ht]
	\centering
        \includegraphics[width=0.8\linewidth]{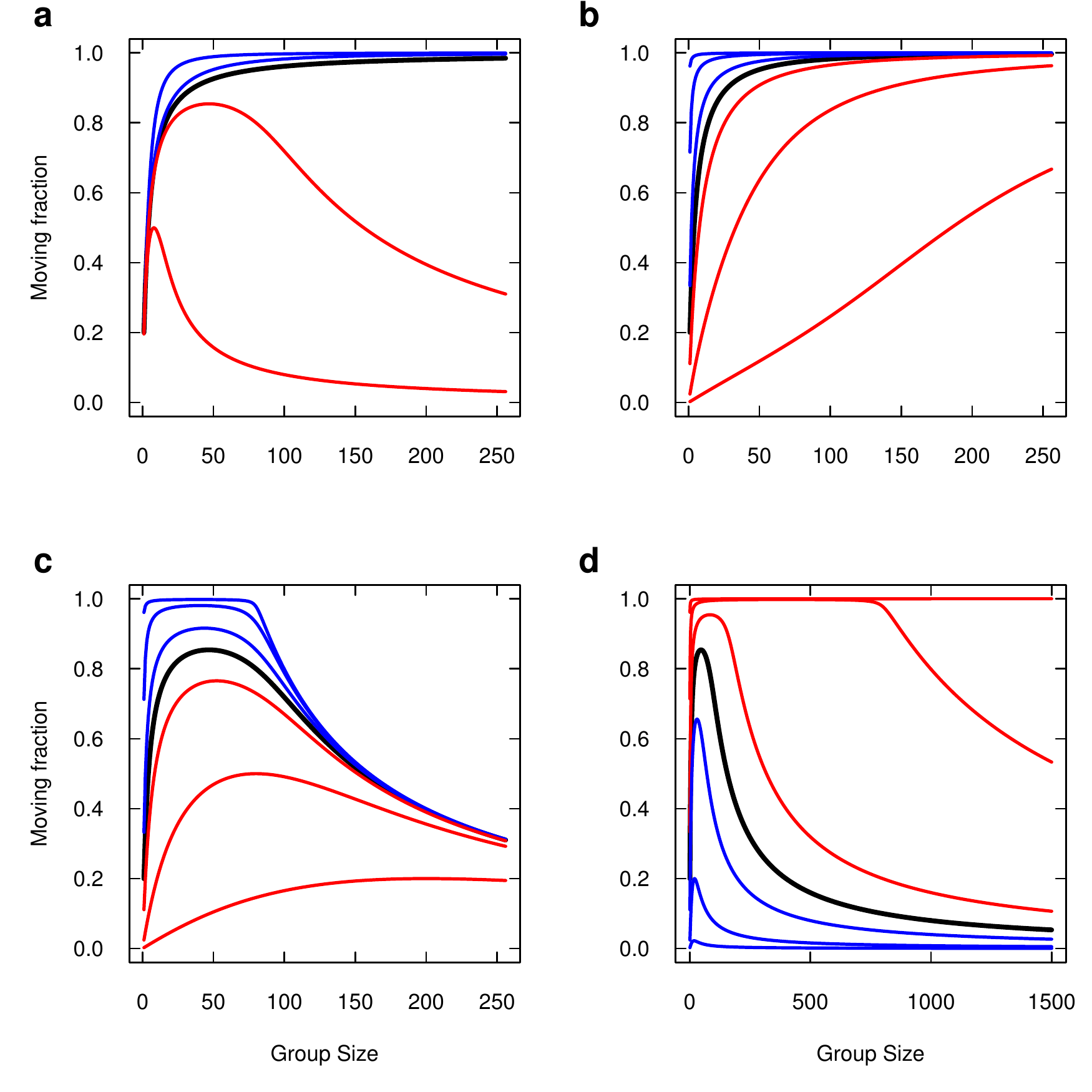}
        \caption{{\bf Moving fraction as a function of group size, varying individual parameters (Variant Model).} 
Same legend as Fig~\ref{fig6} (same parameters, same parameter variations). Note the expanded scale in (d) to clearly show the effect ; the upper red curve starts to decrease around $N=8000$.
}
\label{fig7}
\end{figure}

\begin{figure}[!ht]
	\centering
        \includegraphics[width=1.0\linewidth]{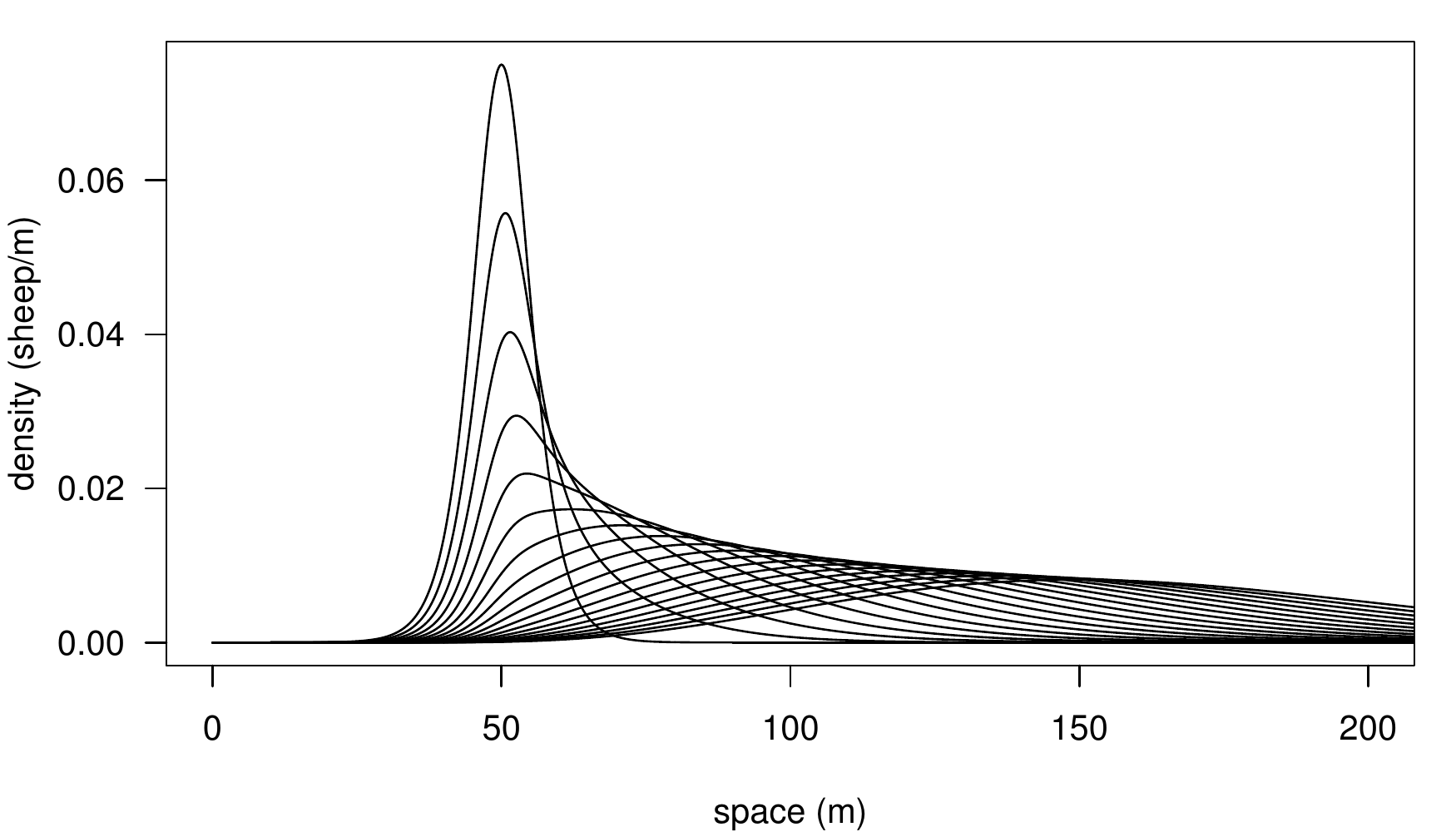}
        \caption{{\bf Propagation with no interaction (1).} 
Numerical simulations of Eqs.~\ref{eq:boltz1}-\ref{eq:boltz2}, with $\Delta t=10^{-2}$ s, and $\alpha_{A}=\alpha_{I}=0$. The leftmost profile is the initial condition, and profile are shown every 100 s. Since there is no non linear term to compensate for diffusion, the profile tends to a gaussian distribution. The center of mass is simply advected towards positive abscissa due to the moving fraction.
}
\label{S1_Fig}
\end{figure}

\begin{figure}[!ht]
	\centering
        \includegraphics[width=0.6\linewidth]{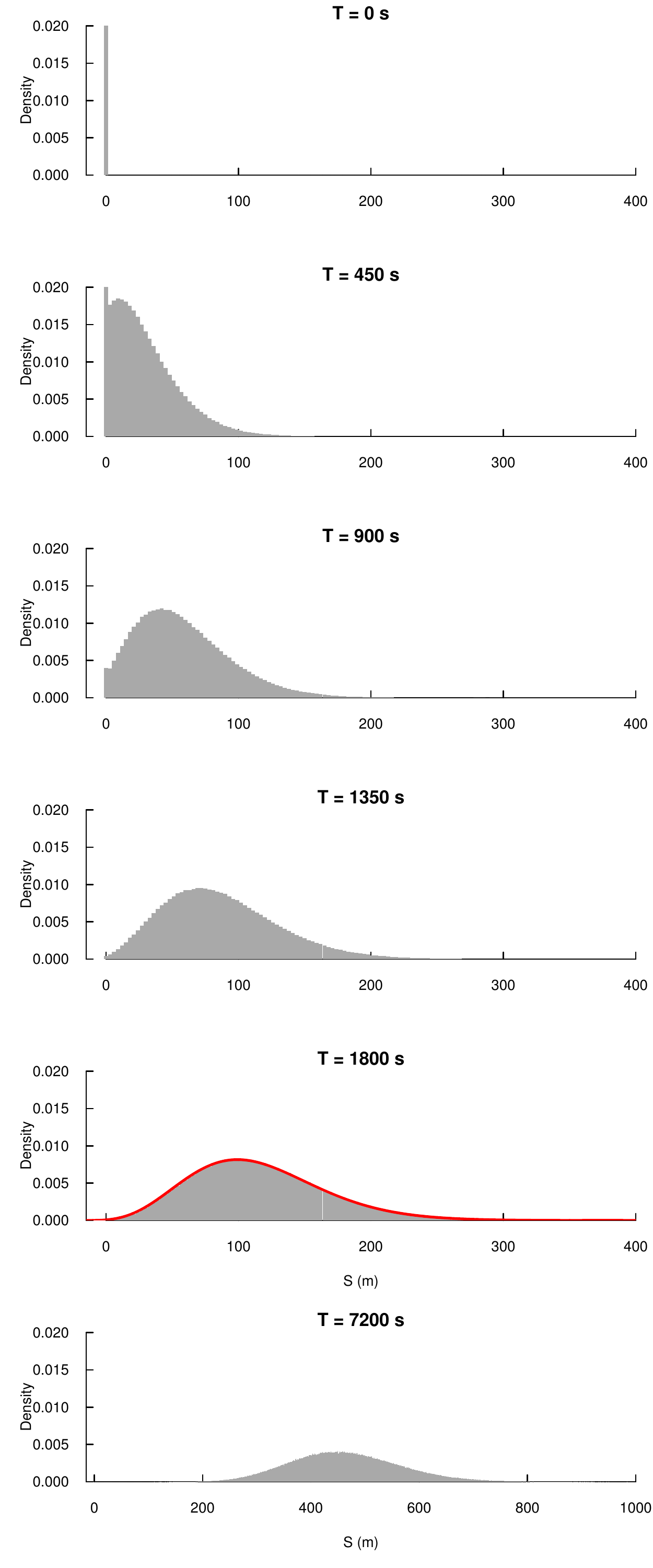}
        \caption{{\bf Propagation with no interaction (2).} 
Statistics over stochastic Monte Carlo simulation of the individual-based model, with $\alpha_{A}=\alpha_{I}=0$. Histograms of presence are reported at different dates, starting from time 0 where all individuals are collapsed near location 0. The corresponding profile from numerical resolution of  Eqs.~\ref{eq:boltz1}-\ref{eq:boltz2} is superimposed at time 1800 s (red curve).
}
\label{S2_Fig}
\end{figure}

Avowedly, extending this property to very large groups would result in a vanishing spontaneous stopping rate, so that the moving fraction should tend to 1 in any case at first sight. 
This is actually true only in the symmetrical case.
In the general case, this effect combines with other parameters and yields non monotonous trends with group size, so we expose it for the sake of interest.
Fig~\ref{fig7} reports the moving fraction as a function of group size (or total mass), varying the parameters the very same way as in Fig.~\ref{fig6}, but for the variant model (graphics can be compared one to one). 

Introducing the variant produces a striking effect upon the sensitivity to parameters.
First, as expected, the moving fraction increases with group size under the symmetrical influences case (Fig~\ref{fig7}a, black curve).
Promoting departure imitation over stopping imitation ($\alpha_{A}>\alpha_{I}$) just accelerate this trend (Fig~\ref{fig7}a, blue curves).
For such promoted departure imitation ($\alpha_{A}>\alpha_{I}$), the spontaneous stopping rate $\mu_{I}$ only affects the rate at which the moving fraction increases with group size (Fig~\ref{fig7}b).

In contrast with the first model, promoting stopping imitation (Fig~\ref{fig7}a, red curves) now displays a non monotonous trend, as there is a range of groups that display an increased moving fraction despite stopping is promoted, and the trend to lower and lower moving fractions is recovered only for largest group (tends to 0 for largest groups). 
In such a case of promoted stopping imitation, this range of group size is strongly affected both by  $\mu_{I}$ (Fig~\ref{fig7}c) and by $\mu_{A}$ (Fig~\ref{fig7}d).
Overall, if stopping imitation is stronger than departure imitation, the variant model promotes the moving fraction in a limited range of group sizes. This moving fraction can be as high as one, and both group size range and maximal value are controlled by the spontaneous rates.

\section*{Discussion}
The analytical solution for the minimal model (a traveling pulse in sech square shape) can be regarded as the solution of reference for collective motion emerging among stop-go entities which synchronise their switching decisions based on a simple behind/ahead partitioning of their influential neighbours.
With no constraint, the dynamics will always converge towards a single pulse since the interaction promote cohesion in any case: individuals ahead waits for those behind to keep up while individuals behind do what it takes to keep up. 
The mechanism should ensure the self-organised convergence to only one traveling pulse from any initial condition, and restore from any perturbation (see \nameref{S7_Movie}).
Incorporating further ingredients could alter its cohesion and its speed. 
It is likely, for instance, that taking back the inhibitory part into account would affect both the propagation speed, and the shape of the traveling pulse.
However, it would still ensure the propagation of the group as a traveling pulse, as indicated by the numerical solution of the complete model using the Boltzmann-like equations.

\bigskip

In the present context, we had no reason to take into account the biological limits to perception, and in the model, the stimulating neighbours are integrated over infinite half-lines. 
Avowedly, animals do have perceptual limits, being they endogenous or due to fragmented landscape.
However, groups were small enough to neglect crowding effect upon perception (which would justify a topological limit on influential neighbours set, like in starling flocks or large fish flocks\cite{Ballerini2008,Rosenthal2015}), and landscape obstacles to perception would need to be introduced explicitly if they were of relevance. 
Moreover, the dynamics favours packing against diffusion, so if small groups start from reasonably dense initial condition, the probability that the group disperse so widely that individuals could not see each other anymore due to endogenous limit is nearly zero. 
In absence of external factors disrupting the groups, introducing a biologically relevant metric cutoff (e.g. some hundreds of meters in sheep) would then have no effect upon the sustained dynamics of the pulse (which is far narrower than that). 

With no limited perception, the spatial effect  results entirely from the asymmetrical influence of individuals that are in the opposite state: only active individuals ahead are stimulating switching to motion, only inactive individuals behind are stimulating switching to stop. 
The simple behind/ahead asymmetrical influence together with the double mimetic effect are sufficient to generate the traveling pulse since it promotes the tendency to wait at the front edge of the pulse, and to keep moving at the back edge \cite{Mogilner1999}.

Without this simple asymmetry, e.g. if we had considered all (behind and ahead) active individuals as stimulating switching to motion, there would be no spatial effect at all, since the stimulation would be the same all over the space. 
In such a case, the dynamics would degenerate into a simple advection-diffusion process, and the group would eventually disperse despite interactions. 
This asymmetry can then be seen as an alternative to models based upon topologically-defined neighbours\cite{Ballerini2008a} or limited sensing kernels (non local terms)\cite{Eftimie2007}.
It could as well be described by an Heaviside odd kernels on the half-line. 
Considering extension to 2-dimensional motion, our simple behind/ahead symmetry breaking parallels the violation of Newton’s third law (action-reaction symmetry) in models based on social forces\cite{Barberis2016}.

\bigskip
Classically, traveling pulse studies start directly from a macroscopic description at the system level \cite{Kerner1994,Purwins2005,Purwins2010}.
In the present study, we have found a traveling pulse solution starting from the ``microscopic'' description of interactions at the individual level (and even binary interactions in the minimal model) so that the macroscopic solution (in the minimal version of the model) is completely parametrized by the individual behavioral parameters.

In the same spirit, Bertin et al. \cite{Bertin2009}, extended by Peshkov et al. \cite{Peshhkov2014}, propose a method to derive density equations for the Vicsek model \cite{Vicsek1995} in the dilute regime (binary interactions).
Starting from the Boltzmann expression (and using an approximation needed by the 2-dimensional nature of their model), they find an explicit expression of the macroscopic transport coefficients. 
Projecting their Boltzmann expression onto an arbitrary direction in the unstable collective motion regime, the resulting 1-dimensional system display irregular trains of traveling pulses.
In contrast to the $\mathrm{sech}^{2}$ profile we found for our model, these traveling pulses profiles are made of two (behind / ahead) exponential decays compatible with their hydrodynamic approximation. 
Saragosti et al.\cite{Saragosti2010} also found double exponential wave profiles by deriving analytical macroscopic behavior from a kinetic description of the mesoscopic run-and-tumble process in chemotactic bacteria \textit{E. coli}.
Such traveling bands have been long identified in large-scale IBM simulations \cite{Chate2008,Ginelli2010}.
The Vicsek model assuming a constant velocity module, it would be interesting to study the effect of incorporating coupled intermittent motion in such large-number 2-dimensional systems, e.g. along the lines developed in \cite{Bertin2017}.

\section*{Methods}

\subsection*{Ethics statement} Animal care and experimental manipulations were applied in conformity with the rules of the Ethics Committee for Animal Experimentation of Federation of Research in Biology of Toulouse, in accordance with the European Directive 2010/63/EU, with the rules of the European Convention for the Protection of Vertebrate Animals used for Experimental and Other Scientific Purposes.
All protocols were approved by the Steering Committee of the National Institute of Higher Education in Agricultural Sciences - Montpellier SupAgro (French Ministry of Agriculture). 
We note that upon the French Ethical Committee for animal experimentation regulation, no special rule had to be invoked since no protected or endangered species was involved, and the experiments did not imply any invasive nor stressful manipulation, the experimental protocol consisting only in the observation of groups and the acquired data being only pictures of the animals in their normal herding conditions. 
At the end of the experiment, all animals reintegrated the herd of the breeding research station.
All personnel involved had technical support and supervision by the employees of the Research Station as required by the French Ministry of Research.

\subsection*{Data collection}
Sheep (Merinos d’Arles) groups evolutions were collected at the experimental farm of Domaine du Merle (5.74$^\circ$E and 48.50$^\circ$N, South France) during 2008-2009 winter. 
Groups of 18-months aged females were formed, picking individuals at random from a large sheep herd (around 1600) which was raised on the domain.
The groups were introduced within one of four 80m x 80m enclosures delimited by fences and opaque 1.2m high polypropylene blind (for visual isolation).
The pastures were flat and homogeneously covered by native Crau grass. 
A 7-m-high tower was anchored at the middle point between enclosures, from the top of which snapshots of groups were recorded every second for an hour, using Digital cameras (15.1-megapixel Canon EOS D50).
Only the second half-hour recording was used in data analysis, to discard perturbation effects due to the introduction of groups in enclosures.
Groups of N = 2, 3, 4 and 8 individuals were used, with 8 replications each.
2 groups of 8 individuals, recorded on the same day, were discarded from the analysis because the high wind condition was very perturbative to their behaviour (they kept about motionless for an hour near the blind that was the most protective from the wind).

\subsection*{Events extraction}
For groups of N = 2, 3 and 4, the position of each sheep was visually tracked using a Cintiq interactive pen displays [Cintiq 21 UXGA 1600 x 1200 pixels). 
From these positional tracks, events of collective departures and collective stops were identified, and we visually checked on the original pictures that they well corresponded to head-up walking behaviours.
For groups of N = 8, harder to track, we first identify such events on the original pictures, and only tracked the position from the start of collective departures to the end of collective stops.
Field coordinates were recovered from pixel coordinates using projective geometry inverse.

Finally, we obtained 76, 58, 66 and 21 collective departures events for groups of N=2, 3, 4 and 8 ; and 73, 56, 60 and 18 collective stop events respectively (the lower number of collective stops is because we filtered out the few events where the initiator stopped before the last follower departed, so that the stimuli at work were not clearly determined). 

\subsection*{Estimation of interaction parameters $\alpha_{\mathrm{\bullet}},\beta_{\mathrm{\bullet}},\gamma_{\mathrm{\bullet}}$}
We follow the same procedure as we used in previous studies.
In each collective departure event, we considered the \emph{following latency (in s)} for each individual (time elapsed between the previous individual switching to walking and the switching time of this individual). 
We then obtained a collection of latencies, each associated with the states of other individuals in the group (namely, $W$ the number already in walking mode, and $R$ the number still at rest).
The corresponding \emph{following rate} $f(R,W)$ (in $s^{-1}$) was then recovered as the inverse of the mean latency before switching when confronted to $R,W$, taking into account the number of individuals at risk. 
Gathering all those rates across the group size 2, 3, 4 and 8, we performed a single regression in the log-domain following:

\begin{verbatim}
mcreg = MCMCregress(log(LatencesDeparts$f) ~  log(LatencesDeparts$W) 
                                             + log(LatencesDeparts$R) );
\end{verbatim}

We used MCMCregress from the R Package MCMCpack in order to obtain distribution-free confidence interval.
The use of standard lm / confint yielded the same results to the second digit.
The output of lm was:
\begin{verbatim}
Multiple R-squared:  0.9455,	Adjusted R-squared:  0.9334 
F-statistic: 85.04 on 2 and 10 DF,  p-value: 5.281e-07
\end{verbatim} 

\bigskip

We performed the same data analysis for the collective stops.
The corresponding output of lm was:
\begin{verbatim}
Multiple R-squared:  0.9108,	Adjusted R-squared:  0.8929 
F-statistic: 51.02 on 2 and 10 DF,  p-value: 5.662e-06
\end{verbatim}

\subsection*{Spontaneous switching rates}.
To estimate the spontaneous rate of switching to the stopped state $\mu_{I}$, we consider the set of all durations between the starts of collective move (the date at which the last individual had switched to the moving state) and the date at which the first moving individual switched to the stopped state. The corresponding rate appeared to depend upon the group size, following
\begin{equation}\label{mun}
\mu_{I} = \mu^{*}_{I}/N
\end{equation}
with $\mu^{*}_{I}= 0.08 \ s^{-1}$.

\bigskip

Unfortunately, we found impracticable to estimate accurately the spontaneous rate of switching to the moving state $\mu_{A}$.
Indeed, the spontaneous departure of one individual could trigger a collective response in some cases, but lots of them actually do not, because the inhibitory effect of the others makes it stop before they start moving.
Such aborted departures would mix with the high number of small moves that sheep display while grazing, when one individual leave the grass clump he was feeding on, walks a couple of steps and resume grazing on another clump.
It was thus impossible to define a clear behavioural clue to cut among pure grazing small moves and actual aborted departures.
This parameter remains then free in the present study, and we provide a realistic value, based on Monte Carlo simulations of the whole process, chaining multiple collective departures / collective stops over 1800 s, and calibrating it by comparing model predictions to the average distances experimental groups ranged over the pasture.

\clearpage

\section*{Appendix 1 : Analytical solution in the steady regime for the minimal model}
\label{S1_Appendix}

Let $A(x,t)$ and $I(x,t)$ denote respectively the density of active (moving) and inactive (stopped) sheep at location $x$ at time $t$. 
They evolve according to:

\begin{equation}
\left\{ \begin{array}{ll}
\partial_{t}I(x,t)&=-K_{A}(x,t)I(x,t)+K_{I}(x,t)A(x,t)\vspace{0.2cm}\\
\partial_{t}A(x,t)+v\partial_{x}A(x,t)&=+K_{A}(x,t)I(x,t)-K_{I}(x,t)A(x,t) 
\end{array} \right.
\label{eq:boltz1}
\end{equation}

where $K_{A}(x,t)$ and $K_{I}(x,t)$ are respectively the conversion rates from stopped-to-moving (activation) and moving-to-stopped (inactivation) at location $x$ at time $t$, which depend on $A$ and $I$ according to:

\begin{equation}
\left\{ \begin{array}{ll}
K_{A}(x,t) &= \mu_{_{A}} 
  + \alpha_{_{A}}{\left[  \int^{\infty}_{x} A(u,t)du  \right]}^{\beta_{A}} {\left[ N - \int^{\infty}_{x} A(u,t)du  \right]}^{-\gamma_{A}} \vspace{0.2cm}\\
K_{I}(x,t) &= \mu_{_{I}} 
  + \alpha_{_{I}}{\left[  \int^{x}_{-\infty} I(u,t)du  \right]}^{\beta_{I}} {\left[ N - \int^{x}_{-\infty} I(u,t)du  \right]}^{-\gamma_{I}} 
 \end{array} \right.
\label{eq:boltz2}
\end{equation}

with $N=\int^{\infty}_{-\infty} A(u,t)+I(u,t)du$ is the total amount of sheep (which is conserved in time), and parameters are those given in the individual-based model.
This description in density is the direct translation of the IBM expressions (in the limit of continuum theory).

Let now consider the classical macroscopic descriptors $\eta(x,t)$, the sum density of sheep (moving and stopped) at location $x$ at time $t$, and the corresponding flow $j(x,t)$, defined by:

\begin{equation}
\left\{ \begin{array}{ll}
\eta(x,t) &= A(x,t) + I(x,t) \vspace{0.2cm}\\
j(x,t) &= v\ A(x,t) + 0\  I(x,t)
\end{array} \right.
\label{eq:macro1}
\end{equation}

Introducing $\beta(x,t)$, the moving fraction at location $x$ at time $t$, defined by:

\begin{equation}
\beta(x,t) = \frac{A(x,t)}{A(x,t) + I(x,t)} = \frac{A(x,t)}{\eta(x,t)}
\label{eq:macro2}
\end{equation}

we have $A(x,t)=\beta(x,t)\eta(x,t)$, so that $j(x,t)=v\ \beta(x,t) \eta(x,t)$.

\bigskip

The macroscopic description in Eq~(\ref{eq:macro1}) can then be equivalently expressed by:

\begin{equation}
\left\{ \begin{array}{ll}
\eta(x,t) &= A(x,t) + I(x,t) \vspace{0.2cm}\\
\beta(x,t) \eta(x,t) &= A(x,t) 
\end{array} \right.
\label{eq:macro3}
\end{equation}

which we will use from now on.
The evolution of these macroscopic descriptors is derived from Eq~(\ref{eq:boltz1}), summing the two evolutions to obtain $\eta$, and using only the second as for $\beta \eta$, leading to:

\begin{equation}
\left\{ \begin{array}{lll}
\partial_{t}\eta &+ v\ \partial_{x}(\beta\eta) &= 0 \vspace{0.2cm}\\
\partial_{t}(\beta\eta) &+ v\ \partial_{x}(\beta\eta) &= K_{A}(1-\beta)\eta - K_{I}\beta\eta 
\end{array} \right.
\label{eq:macro4}
\end{equation}

in which dependencies to $(x,t)$ have been omitted for the sake of clarity, and where $K_{A}$ and $K_{I}$ are now expressed in macroscopic terms, following:

\begin{equation}
\left\{ \begin{array}{lll}
K_{A}(x,t) &= \mu_{A} + \alpha_{A}\ {\left[ A^{+}(x,t)\right]}^{\beta_{A}} \ &{\left[ N - A^{+}(x,t)  \right]}^{-\gamma_{A}} \vspace{0.2cm}\\
K_{I}(x,t) &= \mu_{I} + \alpha_{I}\ {\left[  I^{-}(x,t)  \right]}^{\beta_{I}} \ &{\left[ N - I^{-}(x,t)  \right]}^{-\gamma_{I}}
\end{array} \right.
\label{eq:macro5}
\end{equation}

with $A^{+}(x,t)$ the quantity of moving sheep ahead of $x$, and $I^{-}$ the quantity of stopped sheep behind $x$ :

\begin{equation}
\left\{ \begin{array}{ll}
A^{+}(x,t) &= \int^{\infty}_{x} \beta(u,t)\eta(u,t)du \vspace{0.2cm}\\
I^{-}(x,t) &=  \int^{x}_{-\infty} (1-\beta(u,t))\eta(u,t)du
\end{array} \right.
\label{eq:macro6}
\end{equation}

The complete macroscopic minimal model reads:

\begin{equation}
\left\{ \begin{array}{lll}
\partial_{t}\eta(x,t) &+ v\ \partial_{x}(\beta(x,t)\eta(x,t)) &= 0 \vspace{0.2cm}\\
\partial_{t}(\beta(x,t)\eta(x,t)) &+ v\ \partial_{x}(\beta(x,t)\eta(x,t)) &= \eta(x,t)  \times \\

&& {\Big[} (1-\beta(x,t))(\mu_{A} + \alpha_{A} \int^{\infty}_{x} \beta(u,t)\eta(u,t)du) \vspace{0.2cm}\\ 
&&- \beta(x,t)(\mu_{I} + \alpha_{I}  \int^{x}_{-\infty} (1-\beta(u,t))\eta(u,t)du) \Big] 
\end{array} \right.
\label{eq:macro7}
\end{equation}

summarized by:

\begin{equation}
\left\{ \begin{array}{lll}
\partial_{t}\eta &+ v\ \partial_{x}(\beta\eta) = 0 \vspace{0.2cm}\\
\partial_{t}(\beta\eta) &+ v\ \partial_{x}(\beta\eta) = \eta{\Big[} (1-\beta)K_{A} - \beta K_{I} \Big] 
\end{array} \right.
\label{eq:macro8}
\end{equation}

where $K_{A}$ and $K_{I}$ are :

\begin{equation}
\left\{ \begin{array}{lll}
K_{A}(x,t) &= \mu_{A} + \alpha_{A}\ A^{+}(x,t)\\
K_{I}(x,t) &= \mu_{I} + \alpha_{I}\  I^{-}(x,t)
\end{array} \right.
\label{eq:macro9}
\end{equation}

with $A^{+}(x,t)$ the quantity of moving sheep ahead of $x$, and $I^{-}$ the quantity of stopped sheep behind $x$ :

\begin{equation}
\left\{ \begin{array}{ll}
A^{+}(x,t) &= \int^{\infty}_{x} \beta(u,t)\eta(u,t)du \vspace{0.2cm}\\
I^{-}(x,t) &=  \int^{x}_{-\infty} (1-\beta(u,t))\eta(u,t)du
\end{array} \right.
\label{eq:macro10}
\end{equation}

\bigskip

This expression gives the evolution of the density of sheep in a fixed frame $(x,t)$, attached to the field (hereafter the field frame).

\subsection*{Non Homogeneous Steady State}
Numerical simulations suggest the existence of a non homogeneous steady state with the form of a propagating wave.
Here, we characterize this state.

Having in mind a solution in the form of a wave propagation, we will rewrite the system in another frame $(y,t)$, which is moving at a constant speed $c$ relatively to the field frame, with coordinates:

\begin{equation}
\left\{ \begin{array}{ll}
y &= R(x,t) = x - ct\\
t &=  Q(x,t) = t
\end{array} \right.
\label{eq:rep1}
\end{equation}

with coincidence of the two frames at initial time.
We then have, for any $g(y,t) = f(R(x,t),Q(x,t))$:

\begin{equation}
\left\{ \begin{array}{lll}
\partial_{x}f &=  \partial_{x}R\ \partial_{y}g + \partial_{x}Q\ \partial_{t}g & =  \partial_{y}g\\
\partial_{t}f &= \partial_{t}R\ \partial_{y}g + \partial_{t}Q\ \partial_{t}g     & = \partial_{t}g -c\ \partial_{y}g\\
\end{array} \right.
\label{eq:rep2}
\end{equation}

\bigskip

Denoting $\tilde{\eta}(y,t)=\eta(R(x,t),Q(x,t))$ the density of sheep, and $\tilde{\beta}(y,t)=\beta(R(x,t),Q(x,t))$ the moving fraction in the moving frame, we then have from Eq \ref{eq:macro2}:

\begin{equation}
\left\{ \begin{array}{lll}
\partial_{t}\tilde{\eta} -c\ \partial_{y}\tilde{\eta} &+ v\ \partial_{y}(\tilde{\beta}\tilde{\eta}) = 0 \vspace{0.2cm}\\
\partial_{t}(\tilde{\beta}\tilde{\eta}) -c\ \partial_{y}(\tilde{\beta}\tilde{\eta}) &+ v\ \partial_{y}(\tilde{\beta}\tilde{\eta}) = \tilde{\eta}{\Big[} (1-\tilde{\beta})\widetilde{K}_{A} - \tilde{\beta} \widetilde{K}_{I} \Big] 
\end{array} \right.
\label{eq:mov1}
\end{equation}

where $\widetilde{K}_{A}$ and $\widetilde{K}_{I}$ are:

\begin{equation}
\left\{ \begin{array}{lll}
\widetilde{K}_{A}(y,t) &= \mu_{A} + \alpha_{A}\ \int^{\infty}_{y} \tilde{\beta}(u,t)\tilde{\eta}(u,t)du\\
\widetilde{K}_{I}(y,t) &= \mu_{I} + \alpha_{I}\  \int^{y}_{-\infty} (1-\tilde{\beta}(u,t))\tilde{\eta}(u,t)du
\end{array} \right.
\label{eq:macro3}
\end{equation}

If steady states exist, they obey (in the moving frame):

\begin{equation}
\left\{ \begin{array}{lll}
-c\ \partial_{y}\tilde{\eta} &+ v\ \partial_{y}(\tilde{\beta}\tilde{\eta}) = 0 \vspace{0.2cm}\\
-c\ \partial_{y}(\tilde{\beta}\tilde{\eta}) &+ v\ \partial_{y}(\tilde{\beta}\tilde{\eta}) = \tilde{\eta}{\Big[} (1-\tilde{\beta})\widetilde{K}_{A} - \tilde{\beta} \widetilde{K}_{I} \Big] 
\end{array} \right.
\label{eq:ss1}
\end{equation}

that we rewrite as:

\begin{equation}
\left\{ \begin{array}{lll}
-c\ n' &+ v\ (bn)' = 0 \vspace{0.2cm}\\
-c\ (bn)' &+ v\ (bn)' = n{\Big[} (1-b)(\mu_{A} + \alpha_{A} A^{+}) - b(\mu_{A} + \alpha_{A} I^{-}) \Big] 
\end{array} \right.
\label{eq:ss2}
\end{equation}

where steady states $(n,b)$ are such that $n(y) = \tilde{\eta}(y,t)$ and $b(y)=\tilde{\beta}(y,t)$ $\forall t$, prime denotes regular derivative with respect to $y$, and

\begin{equation}
\left\{ \begin{array}{lll}
A^{+} &=  \int^{\infty}_{y} b(u)n(u)du\\
I^{-} &=  \int^{y}_{-\infty} (1-b(u))n(u)du
\end{array} \right.
\label{eq:ss3}
\end{equation}

\bigskip

\subsection*{Propagation speed}
A pure propagation of a wave would imply a pure advection of a non homogenous profile of $n$.
In the static frame, this would imply that the first equation in Eq.\ref{eq:macro2} is a pure advection, meaning:

\begin{equation}
\partial_{t}\eta + v\ \partial_{x}(\beta\eta) = \partial_{t}\eta + v\ \partial_{x}\eta = 0
\label{eq:nhss1}
\end{equation}

This can occur only when $\partial_{x}\beta = 0$, meaning an homogeneous profile for $\beta$, or equivalently, in the moving frame, $b(y)=b_{s},\ \forall y$.

Plugging into the first equation of Eq.\ref{eq:ss2}, this translates into:

\begin{equation}
-c\ n' + v\ (b_{s}n)' = -c\ n' + v\ b_{s}\ n' = 0 
\label{eq:nhss2}
\end{equation}

yielding

\begin{equation}
c = b_{s}v
\label{eq:nhss3}
\end{equation}

meaning, consistently, that the density profile would propagate at a speed $b_{s}v$ equal to the average speed (the moving fraction times the individual speed).

\subsection*{Wave profile for the density $n(y)$}

What would be the steady state density profile in this particular moving frame?
Second equation in Eq~(\ref{eq:ss2}) becomes:

\begin{equation}
\begin{array}{ll}
vb_{s}(1-b_{s})n' = n \Big[ & (1-b_{s})\left(\mu_{A} + \alpha_{A} b_{s} \int^{\infty}_{y} n(u)du\right) \\
                                         & - b_{s}\left(\mu_{I} + \alpha_{I} (1-b_{s}) \int^{y}_{-\infty}n(u)du\right) \Big] 
\end{array}
\label{eq:steady1}
\end{equation}

Rearranging terms, we get:

\begin{equation}
\begin{array}{ll}
vb_{s}(1-b_{s})n' =& n \left[ (1-b_{s})\mu_{A} - b_{s}\mu_{I} \right]\\
& +   n b_{s} (1-b_{s}) \left[ \alpha_{A}  N^{+}  - \alpha_{I}  N^{-} \right] 
\end{array}
\label{eq:steady2}
\end{equation}

\bigskip

with $N^{+}=\int^{\infty}_{y} n(u)du$ and $N^{-}=\int_{-\infty}^{y} n(u)du$.

\bigskip
Deriving with respect to $y$, we get:

\begin{equation}
\begin{array}{lll}
vb_{s}(1-b_{s})n'' 
& =  & n' \left[ (1-b_{s})\mu_{A} - b_{s}\mu_{I} \right]\\
&  +   & n' b_{s} (1-b_{s}) \left[ \alpha_{A}  N^{+}  - \alpha_{I}  N^{-} \right] \\
&  +   & n b_{s} (1-b_{s}) \left[ \alpha_{A}  (-n)  - \alpha_{I}  (n) \right] \\
& =  & n' \left[ (1-b_{s})\mu_{A} - b_{s}\mu_{I} \right]\\
& +   & n' b_{s} (1-b_{s}) \left[ \alpha_{A}  N^{+}  - \alpha_{I}  N^{-} \right] \\
& -   & n^{2} b_{s} (1-b_{s}) \left[ \alpha_{A}  + \alpha_{I}  \right] 
\end{array}
\label{eq:steady3}
\end{equation}

Multiplying by $n$, we obtain:

\begin{equation}
\begin{array}{lll}
vb_{s}(1-b_{s})n'' n
& =  & n' n\left[ (1-b_{s})\mu_{A} - b_{s}\mu_{I} \right]\\
& +   & n' \large\{ nb_{s} (1-b_{s}) \left[ \alpha_{A}  N^{+}  - \alpha_{I}  N^{-} \right] \large\}\\
& -   & n^{3} b_{s} (1-b_{s}) \left[ \alpha_{A}  + \alpha_{I}  \right] 
\end{array}
\label{eq:steady4}
\end{equation}

\bigskip

From Eq.\ref{eq:steady2}, we extract the term in $\{\}$ in Eq.\ref{eq:steady4} :

\begin{equation}
\begin{array}{ll}
n b_{s} (1-b_{s}) \left[ \alpha_{A}  N^{+}  - \alpha_{I}  N^{-} \right]  
&= vb_{s}(1-b_{s})n' - n \left( (1-b_{s})\mu_{A} - b_{s}\mu_{I} \right)
\end{array}
\label{eq:steady5}
\end{equation}

\bigskip

that we plug back into Eq.\ref{eq:steady4}, and we obtain:

\begin{equation}
\begin{array}{lll}
vb_{s}(1-b_{s})n'' n
& =  & n' n\left[ (1-b_{s})\mu_{A} - b_{s}\mu_{I} \right]\\
& +   & n'  \left[vb_{s}(1-b_{s})n' - n \left( (1-b_{s})\mu_{A} - b_{s}\mu_{I} \right) \right]\\
& -   & n^{3} b_{s} (1-b_{s}) \left[ \alpha_{A}  + \alpha_{I}  \right] \\
& =  & n' n\left[ (1-b_{s})\mu_{A} - b_{s}\mu_{I} \right]\\
& +   & (n')^{2} vb_{s} (1-b_{s}) \\
& -   & n'  n\left[ (1-b_{s})\mu_{A} - b_{s}\mu_{I} \right]\\
& -   & n^{3} b_{s} (1-b_{s}) \left[ \alpha_{A}  + \alpha_{I}  \right] \\
& =  &  (n')^{2} vb_{s} (1-b_{s})\\
& -   & n^{3} b_{s} (1-b_{s}) \left[ \alpha_{A}  + \alpha_{I}  \right]
\end{array}
\label{eq:steady6}
\end{equation}

\bigskip

finally yielding:

\begin{equation}
\begin{array}{lll}
n'^{2} - nn''-\frac{\alpha_{A}  + \alpha_{I}}{v}n^{3}=0
\end{array}
\label{eq:steady7}
\end{equation}

\bigskip

Let $\xi= (\alpha_{A}  + \alpha_{I})/v$.

A solution to Eq.\ref{eq:steady7} compatible with a wave peak  at $y=0$ (i.e. $n'(0)=0$) is:

\begin{equation}
\begin{array}{lll}
n(y)= \frac{K}{2\xi}\mathrm{sech}^{2}\left((1/2) \sqrt{K} y\right)
\end{array}
\label{eq:steady8}
\end{equation}

\bigskip

with $K$ a constant.
To solve for this constant, we consider the constraint that $\int_{-\infty}^{+\infty}n(u)du=\mathrm{N}$, the number of sheep, is conserved.

Since

\begin{equation}
\begin{array}{lll}
\int_{-\infty}^{+\infty}\mathrm{sech}^{2}\left(k y\right)dy = 2/k
\end{array}
\label{eq:steady9}
\end{equation}

\bigskip

we then have:

\begin{equation}
\begin{array}{lll}
\int_{-\infty}^{+\infty}n(u)du = \int_{-\infty}^{+\infty}du\frac{K}{2\xi}\mathrm{sech}^{2}\left((1/2) \sqrt{K} u\right) = \frac{\sqrt{K}}{2\xi}=\mathrm{N}
\end{array}
\label{eq:steady10}
\end{equation}

so that 

\begin{equation}
\begin{array}{lll}
K = (1/4)\ \mathrm{N}^{2}\ \xi^{2}
\end{array}
\label{eq:steady11}
\end{equation}

\bigskip

Finally, the profile for the density, in the frame moving at speed $\beta_{s}v$, is then:

\begin{equation}
\begin{array}{lll}
n(y)= \frac{1}{2}\mathrm{N}\gamma \mathrm{sech}^{2}\left(\gamma y\right)
\end{array}
\label{eq:steady12}
\end{equation}

with 

\begin{equation}
\gamma = \frac{\mathrm{N}(\alpha_{A}+\alpha_{I})}{4v}
\label{eq:steady13}
\end{equation}

\bigskip

We note that this solution $n(y)$ for this steady state \emph{profile} is independent from the value $b_{s}$ as well as from the spontaneous activation/inactivation parameters $\mu_{A}$ and $\mu_{B}$. We see below that those parameters are only involved in the propagation speed of this profile.

\subsection*{Value of the moving fraction $b_{s}$}

The solution above is compatible with the model only if the corresponding moving fraction $b_{s}$ is such that $b_{s} \in [0..1]$.

This solution with a wave peak at $y=0$ is compatible with only one value for $b_{s}$, since we must have that Eq.\ref{eq:steady1} holds at $y=0$, namely :

\begin{equation}
\begin{array}{ll}
vb_{s}(1-b_{s})n'(0) = n(0) \Big[ & (1-b_{s})\left(\mu_{A} + \alpha_{A} b_{s} \int^{\infty}_{0} n(u)du\right) \\
                                                   & - b_{s}\left(\mu_{I} + \alpha_{I} (1-b_{s}) \int^{0}_{-\infty}n(u)du\right) \Big] 
\end{array}
\label{eq:steady14}
\end{equation}

With $n'(0)=0$ and $n(0)>0$, $b_{s}$ must then be solution to:

\begin{equation}
(1-b_{s})\left(\mu_{A} + \alpha_{A} b_{s} \int^{\infty}_{0} n(u)du\right) - b_{s}\left(\mu_{I} + \alpha_{I} (1-b_{s}) \int^{0}_{-\infty}n(u)du\right)=0 
\label{eq:steady16}
\end{equation}

The wave solution being symetrical around $y=0$, we have:

\begin{equation}
\int^{\infty}_{0} n(u)du = \int^{0}_{-\infty}n(u)du = N/2
\label{eq:steady15}
\end{equation}

so that $b_{s}$ must be solution to:

\begin{equation}
(1-b_{s})\left(\mu_{A} + \alpha_{A} b_{s} N/2\right) - b_{s}\left(\mu_{I} + \alpha_{I} (1-b_{s}) N/2\right) = 0
\label{eq:steady16b}
\end{equation}

i.e.
\begin{equation}
b_{s}^{2}\left[N/2(-\alpha_{A}+\alpha_{I})\right] + b_{s}\left[N/2(\alpha_{A}-\alpha_{I}) - (\mu_{A}+\mu_{I})\right]  + \mu_{A}=0
\label{eq:steady17}
\end{equation}

i.e.
\begin{equation}
C\, b_{s}^{2} - \left(C - (\mu_{A}+\mu_{I})\right)\, b_{s}  - \mu_{A}=0
\label{eq:steady17}
\end{equation}

with $C = \frac{N}{2}(\alpha_{A}-\alpha_{I})$.
\bigskip

We note that, in the symetrical case where $\alpha_{A}=\alpha_{I}$, we have $b_{s}=\mu_{A}/(\mu_{A}+\mu_{I})$.

\bigskip
In other cases, we check that :

\begin{equation}
\left( C - \mu_{A}-\mu_{I} \right)^{2} +4\mu_{A}C = \left( C + \mu_{A}-\mu_{I} \right)^{2} +4\mu_{A}\mu_{I} 
\label{eq:steady19}
\end{equation}

which is always positive because parameter rates are positive or null ; hence solutions to Eq.\ref{eq:steady17} are real.

\bigskip

The correct solution is actually always given by the root:

\begin{equation}
b_{s} = \frac{\left( C - \mu_{A}-\mu_{I} \right) +\sqrt{\left( C - \mu_{A}-\mu_{I} \right)^{2} +4\mu_{A}C}}{2C}
\label{eq:steady18}
\end{equation}

\bigskip

Of course, $C$ can be any real value (negative or positive), depending on $\alpha_{A}-\alpha_{I}$.
However, we have: 
\begin{equation}
\begin{array}{lll}
\lim\limits_{C \rightarrow -\infty} b_{s}(C) &=& 0\\
\lim\limits_{C \rightarrow +\infty} b_{s}(C) &=& 1\\
b_{s}'  (C) &>& 0\; \forall C
\end{array}
\label{eq:steady20}
\end{equation}

so that there always is a solution $b_{s} \in [0..1]$ for any set of parameters (see Fig. S3 for some illustration).

\subsection*{Complete solution in the field frame}

A non homogeneous steady regime is then a propagating wave with a spatial profile in $\mathrm{sech}^{2}$ with:

\bigskip

\begin{equation}
\left\{ \begin{array}{lll}
\eta^{s}(x,t) &= \frac{\mathrm{N}}{2}\frac{\mathrm{N}(\alpha_{A}+\alpha_{I})}{4v} \mathrm{sech}^{2}\left(\frac{\mathrm{N}(\alpha_{A}+\alpha_{I})}{4v} \ (x-b_{s}^{*}vt)\right)\vspace{0.3cm}\\

\beta^{s}(x,t) &= b_{s}^{*} = \frac{\left( \frac{N}{2}(\alpha_{A}-\alpha_{I}) - \mu_{A}-\mu_{I} \right) +\sqrt{\left( \frac{N}{2}(\alpha_{A}-\alpha_{I}) - \mu_{A}-\mu_{I} \right)^{2} +4\mu_{A}\frac{N}{2}(\alpha_{A}-\alpha_{I})}}{N(\alpha_{A}-\alpha_{I})}
\end{array} \right.
\label{eq:appsteady21}
\end{equation}

\bigskip
when $\alpha_{A}\neq\alpha_{I}$.

In the symetrical case where $\alpha_{A}=\alpha_{I}$, the steady state solution simplifies to:

\bigskip

\begin{equation}
\left\{ \begin{array}{lll}
\eta^{s}(x,t) &= \frac{N}{2}\frac{N\alpha}{2v} \ {\mathrm{sech}}^{2}(\frac{N\alpha}{2v} \ (x-b_{s}^{*}vt))\vspace{0.3cm}\\

\beta^{s}(x,t) &= b_{s}^{*} = \mu_{A}/(\mu_{A}+\mu_{I})
\end{array} \right.
\label{eq:appsteady22}
\end{equation}

\clearpage

\section*{Supplemental figures}

\clearpage

\clearpage

\begin{figure}[!ht]
	\centering
        \includegraphics[width=0.6\linewidth]{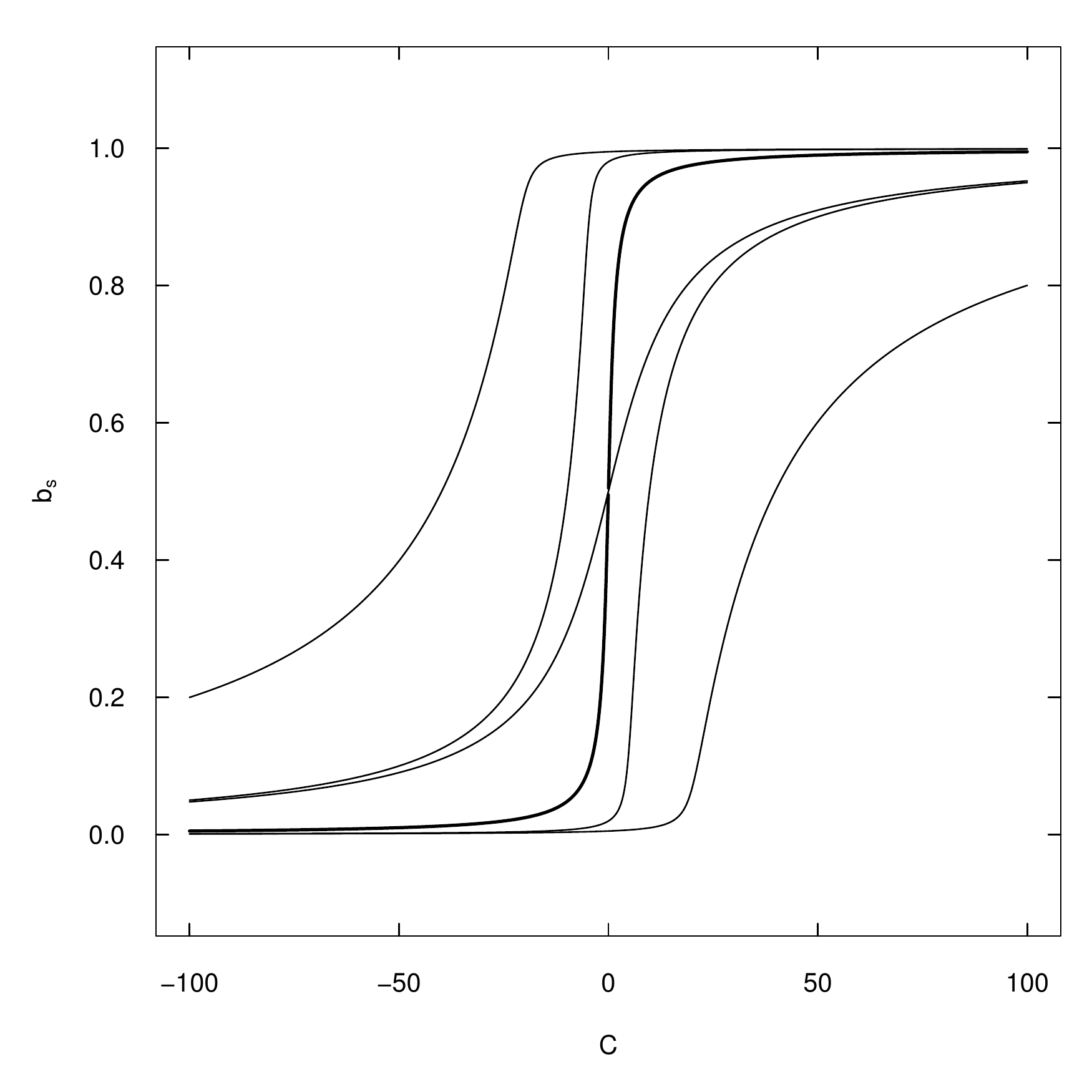}
        \caption{{\bf $b_{s}$ as a function of $C$.} 
The value found by Eq \ref{eq:steady18} in \nameref{S1_Appendix} is given as a function of $C$, and for varied values for $\mu_{A}$ and $\mu_{I}$. Larger values of $C$, meaning $\alpha_{A}>\alpha_{I}$, promote higher moving fractions.
}
\label{S3_Fig}
\end{figure}

\clearpage

\paragraph*{Movie S1 }
\label{S1_Movie}
{\bf Typical motion behaviour of a group of 3 sheep.} 
The evolution of 3 sheep is reported every second in field coordinates. Each sheep has been assigned a colour.
Individuals are nearly motionless most of the time, while they are devoted to grazing. These grazing phases are separated by collective moves that translocate the group over several meters at high speed. A clear event of such a collective motion happens from frame 318 to frame 344.

\paragraph*{Movie S2 }
\label{S2_Movie}
{\bf Stochastic simulations for a group of N=4 sheep with biological parameters given in the main text.} 
One realisation is given in each frame. The 1D position of each individual is presented as a function of time. Horizontal progression indicates a motionless individual. Oblique progression indicates a moving individual. The aggregation in time of oblique  events indicates the synchronisation of motion phases.

\paragraph*{Movie S3 }
\label{S3_Movie}
{\bf Stochastic simulations for a group of N=4 sheep, like in Movie S2, but with a dominant spontaneous departure parameter.} 
The coupling parameters have been downscaled to $\alpha_{A}=0.001$ and $\alpha_{I}=0.016$. As a consequence, individual are mainly driven by independent switching decisions, which results in the loss of synchronisation and group dispersion over ten of meters.

\paragraph*{Movie S4 }
\label{S4_Movie}
{\bf Illustration of the propagation of a group of N=32 sheep predicted by the model}, illustrating that groups can be seen to progress as a whole at some fraction of the individual speed whilst each individual is either stopped or moving at full speed. Stopped individuals are reported by black dots, and moving individuals are reported by red dots. The panel above reports the number of individuals in motion. The dotted line indicates the average of this number over time. 

\paragraph*{Movie S5 }
\label{S5_Movie}
{\bf Numerical simulations of Eqs.~\ref{eq:boltz1}-\ref{eq:boltz2}, with $\Delta t=10^{-2}$ s, for various initial conditions.}
Four different initial conditions were tested: starting at frame 1, the group starts loosely dispersed over 20-30 m with all individuals in the stopped state, from frame 116 the group starts with the same dispersion but with all individuals in the moving state, from frame 210 the group is split into two separated groups with moving individuals in the group ahead, and from frame 348 with moving individuals in the group behind. All simulations converge to the same traveling pulse.

\paragraph*{Movie S6 }
\label{S6_Movie}
{\bf Numerical simulations of Eqs.~\ref{eq:boltz1}-\ref{eq:boltz2}, with $\Delta t=10^{-2}$ s, with analytical solution superimposed.}
The upper panel shows the numerical simulations of the Boltzmann-like equations using the minimal model ($\beta_{A}=\beta_{I}=1$ and $\gamma_{A}=\gamma_{I}=0$, $N=4$, spontaneous rates unchanged). Two sets of parameters are reported: $\alpha_{A}=\alpha_{I}=0.5$ from frame 1, and $\alpha_{A}=0.8,\alpha_{I}=0.2$ from frame 202. The lower panel shows a zoom of the numerical profile, centred on the center of mass (black curve) and the analytical traveling pulse (red curve). Since $\alpha_{A}+\alpha_{I}$ remains equal to 1 in both case, the analytical solution is the same for the density profile. However, the propagating speed is affected (it is faster for the case with dominant activating stimulation) and the route to converge towards the analytical solution is different.

\paragraph*{Movie S7 }
\label{S7_Movie}
{\bf Recovering the traveling pulse propagation after a perturbation.}
We use the IBM simulation program to test how the group react to a perturbation.
Here, the perturbation is the extinction of interactions for a given period of time.
The group starts unperturbed. Stopped individuals are reported in black and moving individuals in red.
In the beginning, we set the camera in constant speed motion tuned to the average speed of the group. 
We can see the group ahead of sync or behind of sync in regards to this moving camera frame, but still it keeps progressing on average. 
At time 1000 (frame 1000), the interactions are set off, and the camera is stopped while its angle is enlarged to cover a larger area. 
From that time, individuals progress at their own pace, leading to group dispersal (by advection/diffusion).
At time 1999 (frame 1999), interactions are restored.
The groups then tends to regain its cohesion, illustrating that the ones ahead waits for the one behind to progress before they move again.
At times near 2500, the group has reached its steady regime density and recovers the steady regime propagation.
The camera is set back in motion at time 2550 and its angle restored to its initial value.

\section*{Acknowledgments}
This work was supported by French Agence Nationale de la Recherche (ANR) Grant BLAN07-3 200418.

\nolinenumbers

%
%
%
\bibliography{P17bib}

\end{document}